\documentclass{article}

\usepackage[numbers,sort&compress]{natbib}
\usepackage{amssymb}
\usepackage{amsmath}
\usepackage{lineno,hyperref}
\usepackage{tablefootnote}
\modulolinenumbers[5]
\usepackage{url}

\usepackage{natbib}
\usepackage{booktabs}
\usepackage{subcaption}
\usepackage{multirow}
\usepackage{float}
\usepackage{tikz}
\usetikzlibrary{shapes.geometric, arrows}
\usepackage{relsize}

\newtheorem{teo}{Theorem}
\newtheorem{lema}{Lemma}
\newtheorem{remark}{Remark}
\def\Proof{{\noindent \bf Proof.\enspace}}
\def\Endproof{\indent }

\title{Model for prognostic of symptomatic, asymptomatic and hospitalized COVID-19 cases with correct demography evolution}
\centering
\author{Antonio Rafael Selva Casta\~neda$^1$,  Erick Eduardo Ramirez-Torres$^2$, \\Luis Eugenio Vald\'es-Garc\'ia$ ^3 $, Hilda Mar\'ia Morandeira-Padr\'on$ ^4 $, \\Diana Sedal Yanez$ ^5 $, Juan I. Montijano$ ^6 $ \\and Luis Enrique Bergues Cabrales$ ^7 $\\ 
\small $ ^{1,2,6} $Instituto Universitario de Investigaci\'on de Matem\'aticas y Aplicaciones,\\ \small Universidad de Zaragoza, Zaragoza, Spain\\
\small$ ^3 $ Centro provincial de Higiene y Epidemiolog\'ia, Santiago de Cuba, Cuba e \\ \small Direcci\'on Provincial de Salud p\'ublica, Santiago de Cuba, Cuba\\
\small $ ^4 $ Dirección Provincial de Salud Pública, Santiago de Cuba, Cuba\\
\small $ ^5 $ Facultad de Ciencias Sociales, Universidad de Oriente, Santiago de Cuba, Cuba\\
\small $ ^7 $ Direcci\'{o}n de Ciencia e Innovaci\'{o}n Tecnol\'{o}gica, Centro Nacional de Electromagnetismo Aplicado, \\ \small Universidad de Oriente, Santiago de Cuba, Cuba
  }
\date{}
\begin{document}
\maketitle

\abstract{The aim of this study is to propose a modified Susceptible-Exposed-Infectious-Removed (SEIR) model that describes the behaviour of symptomatic, asymptomatic and hospitalized patients of COVID-19 epidemic, including the effect of demographic variation of population. It is shown that considering a population growth proportional to the total population leads to solutions with a qualitative behaviour different from the behaviour obtained in many studies, where constant growth ratio is assumed. An exhaustive theoretical study is carried out and the basic reproduction number $R_0$ is computed from the model equations. It is proved that if $R_0<1$ then the disease-free manifold is globally asymptotically stable, that is, the epidemics remits.  Global and local stability of the equilibrium points is also studied. Numerical simulations are used to show the agreement between numerical results and theoretical properties. The model is fitted to experimental data corresponding to the pandemic evolution in the Rep\'ublica  de Cuba, showing a proper behaviour of infected cases which let us think that can provide a correct estimation of asymptomatic cases. In conclusion, the model seems to be an adequate tool for the study and control of infectious diseases in particular the COVID-19 disease transmission.}

\section{Introduction}\label{intro}
The epidemiological reports of COVID-19 epidemic in several countries show that this disease has a high transmission rate. Infected cases have been reported in 222 countries \cite{WHO}. Additionally, the COVID-19 epidemic characterizes by the large number of pre-symptomatic and asymptomatic patients \cite{li2020}. This makes difficult to take effective governmental measures that control this disease \cite{mckibbin2021}. These measures depend on the number of infected cases. An efficient tool to estimate this number of cases is the mathematical modelling.

Different mathematical models have been reported to study this disease (see e.g \citep{Estrada2020} or \cite{Saldana2021}). In different studies a constant population is assumed and neither new individuals (births) nor deads are considered as in \cite{Brugnano2021} where a multiregion model is considered to cope with the in-time and in-space heterogeneity of an epidemic or \cite{Neves2020}, where the probability that a infected people is fully asymptomatic is calculated. The effect of demography is not included in these models. In others it is included the variation of the population due to COVID-19 deads, but no births or natural deads are considered. This is the case for example in \cite{Ndairou2020} where the transmissibility of  superspreaders individuals is studied, or in \cite{peng2020} where a model is implemented to predict the inflection point and possible ending time. In \cite{Lin2020}, an individual reaction and governmental action based on some parameters of 1918 influenza pandemic is investigated. In this study, natural deads are considered but no new individuals are introduced.

The assumption that the total population is constant and neither births nor deads are considered is reasonable when modelling epidemics were the disease spreads quickly in the population and dies out within a short time. However, when the population growth or decrease is significant or the disease cause enough deads, the assumption of constant population is not realistic. One of the most usual ways to include births and natural deads is to assume that they are proportional to the total population. If $N(t)$ is the population, $\Lambda$ is the birth ratio and $\mu$ is the dead ratio, then

\begin{equation}
\notag
N'=\left(\Lambda-\mu\right)N.
\end{equation}

If $\Lambda-\mu>0$ the population grows exponentially, if $\Lambda-\mu<0$, the population decreases exponentially and if $\Lambda=\mu$ the population stays constant. If $N(0)=0$ there can not be the new individuals and the population stays zero forever (it is an equilibrium point).

Even though this is the most accepted model for population evolution, many models in epidemics assume a constant ratio for births. This is the case for example in \cite{Ottaviano2022} where a theoretical study of a SAIRS compartmental model with vaccination is made. In \cite{Hu2012} the dynamical behaviour of SIRS epidemic models were studied. In \cite{Saha2020} the individual behavioural response due to information regarding proper precaution is analized, in \cite{Yang2020} a model for evaluating a possibility to prevent, or delay, the local outbreaks of COVID-19 through restriction on travel form is implemented, in \cite{Shah2021} it is analyzed how immigration, protection, death rate, exposure, cure rate and interaction of infected people with healthy people affect the population and in \cite{Shoaib2021} it is proposed a hybrid computation technique to construct an epidemic model. These papers are a few of many others were it is assumed that the population evolves (in absence of infection) according to the equation

\begin{equation}
\notag
N'=\Lambda-\mu N,
\end{equation}
that has a solution $N(t)=\left[N(0)-\frac{\Lambda}{\mu}\right]e^{-\mu t}+\frac{\Lambda}{\mu}$. With this model, if $N(0)<\Lambda/\mu$, the population always grows up (even if $N(0)=0$) to the maximum value $\Lambda/\mu$. This is not realistic if for example the birth ratio is smaller than the dead ratio.

In this paper we consider a model for the evolution of the epidemics that includes terms for the demography evolution (births and deads) in which the ratio of new individuals is proportional to the total population. This amounts relevant differences in the behaviour of the solution with respect to models with constant growth ratio. Thus in this last case, it is usual the existence of a unique disease free equilibrium point and the origin is not an equilibrium point, whereas in our proposal, there can be a manifold of disease-free equilibrium points and the origin is always an equilibrium point.

A predictor of epidemic potential suggested widely in any epidemic (e.g., COVID-19 disease) is the basic reproductive number ($R_0$), which is dimensionless. Definition of $R_0$ is the expected  number of cases directly generated by one case in a population were all individual are susceptible to infection. This definition  considers that no other individuals are infected or immunized  (naturally or through vaccination) \citep{Fraser2009}. This concept has been used in COVID-19 disease \citep{Sy2021}. When $R_0 > 1$ the COVID-19 disease spreads (that is, SARCoV-2 will continue its propagation among susceptibles organisms). On the contrary, $R_0 < 1$ predicts the end of the COVID-19 disease.

Values of $R_0$ are usually estimated from mathematical models, and the estimated value depends on the model used and other parameter values.  Besides, $R_0$ requires be interpreted in the context of each of these models \citep{ramireztorres2021}. In this paper, in addition to a deep theoretical study of the proposed model we obtain the basic reproductive number and show how the behaviour of the solutions depends on the value of $R_0$. The model is tested with data corresponding to the evolution of the epidemic in the Rep\'ublica de Cuba.

In section 2,  we present the mathematical model and analyze it theoretically.  We prove the  positivity of the solutions. Next,  we obtain the basic reproductive number $R_0$ and we prove that the disease free manifold is globally asymptotically stable if  $R_0<1$.  Then, we analyze the stability of the equilibrium points of the system, showing its dependence with the value of $R_0$.  In section 3, we present some numerical simulations that show the concordance between numerical and theoretical results. In section 4, we fit the parameters of the model with real data from Cuba showing that this model can reproduce the time behaviour of symptomatic and hospitalized cases. Then, we do the discussions of results and finally we present the conclusions.

\section{Methods}
\subsection{Mathematical model}

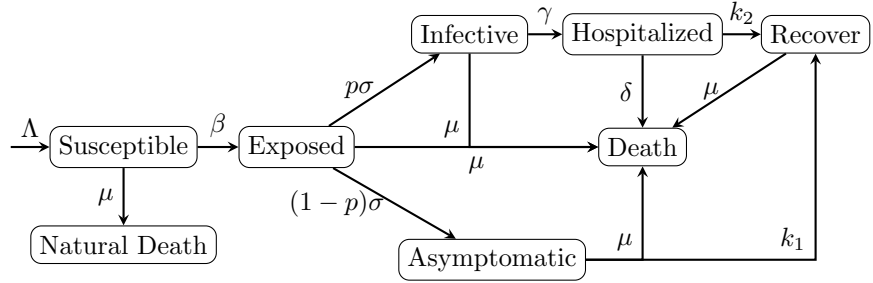
\begin{figure}[h]
\centering
\tikzstyle{startstop} = [rectangle, rounded corners, minimum width=1cm, minimum height=0.5cm,text centered, draw=black]
\tikzstyle{arrow} = [thick,->,>=stealth]
\tikzset{fontscale/.style = {font=\relsize{#1}}}
\begin{tikzpicture}[node distance=1.3cm]
\node (Susceptible) [startstop,fontscale=0.01] {Susceptible};
\node (natud) [startstop, below of=Susceptible,fontscale=0.01] {Natural Death};
\node (expo) [startstop, right of=Susceptible, xshift=1cm,fontscale=0.01] {Exposed};
\node (infec) [startstop, right of=expo, xshift=1cm, yshift=1.5cm,fontscale=0.01] {Infective};
\node (hosp) [startstop, right of=infec, xshift=1cm,fontscale=0.01] {Hospitalized};
\node (death) [startstop, right of=hosp, yshift=-1.5cm,xshift=-1.3cm, ,fontscale=0.01] {Death};
\node (reco) [startstop, right of=hosp, xshift=1cm,fontscale=0.01] {Recover};
\node (asin) [startstop, right of=expo, xshift=1.3cm, yshift=-1.5cm,fontscale=0.01] {Asymptomatic};
\draw [arrow] (Susceptible) -- node [left] {$\mu$} (natud);
\draw [arrow] (-1.5,0)--node [above] {$\Lambda$}(Susceptible) ;
\draw [arrow] (Susceptible) -- node [above] {$\beta$}(expo);
\draw [arrow] (expo) -- node [left] {$p\sigma$}(infec);
\draw [arrow] (expo) -- node [left] {$(1-p)\sigma$}(asin);
\draw [arrow] (infec) -- node [above] {$\gamma$}(hosp);
\draw [arrow] (hosp) -- node [above] {$k_2$}(reco);
\draw [arrow] (hosp) -- node [left] {$\delta$}(death);
\draw [arrow] (asin) -| node [above left] {$k_1$}(reco);
\draw [arrow] (reco) -- node [ left] {$\mu$} (death);
\draw [arrow] (expo) -- node [anchor=north] {$\mu$}(death);
\draw [arrow] (asin) -| node [above left] {$\mu$}(death);
\draw [arrow] (infec) |- node [above left] {$\mu$}(death);
\end{tikzpicture}
\caption{Schematic representation for the formulation of the mathematical model for COVID-19 epidemic.}
\label{my:figf1}
\end{figure}

To simulate the behaviour of the COVID-19 epidemic, we consider an extended  SEIR model with six states: susceptible ($S(t)$), exposed ($E(t)$), infected symptomatic ($I(t)$), infected asymptomatic ($A(t)$), hospitalized-isolated ($H(t)$) and recovered ($R(t)$) as shown in Fig. \ref{my:figf1}.

In the model, $\beta$ (in days$^{-1}$) is the transmission rate, $\mu$ (in days$^{-1}$) the natural death rate, $\Lambda$  (in days$^{-1}$) the birth rate, $ \sigma ^ {- 1} $ (in days) the latent time, $ \gamma ^ {- 1} $ (in days) the time between symptom onset to hospitalization, $p$ the ratio between symptomatic and asymptomatic patients, $k_1^{-1}$ (in days)  the average recovery time of asymptomatic patients, $ k^{-1}_2 $ (in days) the average recovery time of hospitalized patients and $\delta^{-1}$ (in days) is the average time that an hospitalized patient dies. Ramírez-Torres et al. \citep{ramireztorres2021} report that $\beta$ changes over time and depends on transmission rate of SARCoV-2 strain, government measures, risk perception and social responsibility. For simplicity, we will assume that  $\beta$ is constant.

The relationships between these states are represented by the following system of ordinary differential equations
\begin{equation}
\begin{cases} \label{my:eqf1}
\vspace{2mm}
\cfrac{\hbox{d}S}{\hbox{d}t}=\Lambda N-\cfrac{\beta S(I+A)}{N} - \mu S \cr
\vspace{2mm}
\cfrac{\hbox{d}E}{\hbox{d}t}=\cfrac{\beta S(I+A)}{N} - (\sigma + \mu)E \cr
\vspace{2mm}
\cfrac{\hbox{d}I}{\hbox{d}t}=p\sigma E - (\gamma + \mu)I \cr
\vspace{2mm}
\cfrac{\hbox{d}A}{\hbox{d}t}=(1-p)\sigma E - (k_1+ \mu) A \cr
\vspace{2mm}
\cfrac{\hbox{d}H}{\hbox{d}t}=\gamma I - k_2 H -(\delta+\mu) H\cr
\vspace{2mm}
\cfrac{\hbox{d}R}{\hbox{d}t}= k_1 A + k_2 H - \mu R
\end{cases}
\end{equation}
where $N$ is the total population, $N(t)=S(t)+E(t)+I(t)+A(t)+H(t)+R(t)$.  Let us note that in this model, since we are considering births and deads, the total population $N$ changes with time.

We have not included in system \eqref{my:eqf1} the evolution of the dead people since the analysis of the system can be done without it. Nevertheless, we can easily obtain the evolution of the dead people by adding the equation $D'=\mu N+\delta H$.

As mentioned in the introduction, unlike other models \citep{Shoaib2021, Saldana2021, Shah2021, Hu2012, Ottaviano2022, Saha2020, Yang2020, Estrada2020} where the ratio of new individuals is considered constant, we are considering it as $\Lambda N$, proportional to the total population, in agreement with the usual models of population growth, that can be more realistic. We are not including a term representing a limit of the maximum population (carrying capacity) such as in the logistic or the Gompertz equations because we are assuming that the population is far from the carrying capacity and it can not approach the maximum value during the effect of the epidemic. Moreover, most of the results in the paper are also valid if for example a term $-(\Lambda-\mu)\frac{N^2}{N_{\max}}$ is added to the first equation in system \eqref{my:eqf1}.

Denoting $P=(S,E,I,A,H,R)^T$, the system can be formulated in compact form as
\[
\cfrac{\hbox{d} P}{\hbox{d}t} = F(P).
\]
The vector field $F(P)$ is well defined whenever $N\ge 0$. Moreover, variables of the differential system (\ref{my:eqf1}) have biological interpretation only when all of them are non negative. Thus, we will consider the domain
\begin{equation}\label{domain}
\mathcal{D}= \{P\in \mathbb{R}^6\  \hbox{ such that } S, E, I, A, H, R\ge 0\}.
\end{equation}
If $P\in \mathcal{D}$, we will write $P\ge 0$.

\begin{remark}
The proposed system of differential equations has the following properties:
\begin{itemize}
\item
The differential system is autonomous and we can assume, without loss of generality, that the initial time is $t_0=0$.
\item
The vector field is an homogeneous function of $P$, $F(cP)= cF(P)$ for all $c \in R$, and therefore if P(t) is
a solution of the system (with initial conditions $P(0)$) then $cP(t)$ is also a solution of the system (with initial conditions $cP(0)$).
\item
If $P$ is an equilibrium point, then $cP$ is also an equilibrium point for any non zero $c$.
\end{itemize}
\end{remark}

\subsection{Positivity of the solution}

\begin{teo}\label{teo1}
The proposed system of differential equations \eqref{my:eqf1} has a unique solution in $\mathcal{D}$. Moreover, if the initial condition of the system is such that all the variables are non-negative, $P(0) \in \mathcal{D}$, then all variables remain non negative over time. In other words, if the initial condition is in the positive hyperquadrant, the solution remains in this hyperquadrant.
\end{teo} 
\Proof
The vector field $F$ is continuous in $\mathcal{D}$, and also differentiable (except at the zero point). The partial derivatives are uniformly bounded and therefore, the vector field is Lipschitz in the domain $\mathcal{D}$. This guarantees the existence and uniqueness of solution.

Now, to prove the positivity of the solutions, let us suppose that all variables are non negative at a certain time $ t ^ * $. Obviously, $ N (t ^ *)> 0 $ since otherwise we would be in the null equilibrium solution. We will show that if at $ t ^ * $ a variable, let us name it generically $ X (t^*)$, becomes zero the first derivative that does not vanish at $ t ^ * $ must be positive, $ X^{(k)}(t^*)> 0 $ and therefore $ X(t)>0 $ for $ t\gtrsim t^*$.

First, if the $ S $ variable vanishes at $ t^*$ and $N(t^*)>0$, then $S'(t^*)>0 $ and necessarily $S(t)$ is positive for $t \geq t^*$.

It is observed that if $ E(t^*) =I(t^*)=A(t^*)=0 $, their derivatives also vanish at that point and therefore, it would be $ E (t)=I(t)=A(t)=0 $ for all $ t $ (equilibrium point of this subsystem, regardless of the values that $ S, H, R $ have). Furthermore, in this case, $ H'=-(\delta + k_2 + \mu)H$ and $R'= k_2 H - \mu R$, consequently both are non negative for all $ t>t^*$.

If $E(t^*)=0$, then
\[
\begin{array}{l}
E'(t^*)=\beta\dfrac{ S(t^*) (I(t^*)+A(t^*))}{N(t^*)}-(\sigma+\mu)E(t^*),
\\[7pt]
E''(t^*)=S'(t^*)\beta\dfrac{(I(t^*)+A(t^*))}{N(t^*)}+(I'(t^*)
+A'(t^*))\beta\dfrac{ S(t^*)}{N(t^*)}\\
\hspace{9mm}-\beta  S(t^*)(I(t^*)+A(t^*))\dfrac{ N'(t^*)}{N(t^*)^2}-
(\sigma+\mu)E'(t^*).
\end{array}
\]

If $S(t^*) (I(t^*)+A(t^*))>0$, then $E'(t^*)>0$. If $S(t^*)=0$ but $(I(t^*)+A(t^*))>0$ then $E'(t^*)=0$, $S'(t^*)>0$ and $E''(t^*)>0$. If $(I(t^*)+A(t^*))=0$, then $E(t^*)=I(t^*)=A(t^*)=0$. In conclusion, $E(t)>0$ for $ t \gtrsim t^*$.

If this analysis is performed with the other variables and proceeding in a similar way to the $E(t^*)$ case, it is proved that these variables remain non negative at $t\geq t^*$ if they cancel out at $t^*$.
\Endproof

\subsection{Stability of disease free manifold}\label{lastref}

If $E=I=A=0$, the system \eqref{my:eqf1} reduces to

\[
\begin{cases}
\vspace{2mm}
\cfrac{\hbox{d}S}{\hbox{d}t}=\Lambda (S+R+H) - \mu S, \\
\vspace{2mm}
\cfrac{\hbox{d}H}{\hbox{d}t}= - k_2 H -(\delta+\mu) H,\\
\vspace{2mm}
\cfrac{\hbox{d}R}{\hbox{d}t}=k_2 H - \mu R,
\end{cases}
\]
whose general solution is

\[
\begin{array}{lrl}
H(t)=H(0) e^{-(\delta+k_2+\mu)t}, \\
R(t)=-
\dfrac{\delta H(0)}{\delta+k_2}e^{-(\delta+k_2+\mu) t} +(R(0)+\dfrac{\delta H(0)}{\delta+k_2})e^{-\mu t},\\[8pt]
S(t)=-
\dfrac{\Lambda H(0)}{\delta+k_2+\Lambda}\Big(
\dfrac{\delta }{\delta+k_2}+1\Big)e^{-(\delta+k_2+\mu) t}-(R(0)+\dfrac{\delta H(0)}{\delta+k_2})e^{-\mu t}\\ [8pt]
 \Big(S(0)+ R(0)+\dfrac{\delta H(0)}{\delta+k_2}+
\dfrac{\Lambda H(0)}{\delta+k_2+\Lambda}\Big(
\dfrac{\delta }{\delta+k_2}+1\Big)\Big)e^{(\Lambda-\mu)t}.
\end{array}
\]

The solution has three components, all with exponential behaviour, $e^{-(\delta+k_2+\mu)t}$, $e^{-\mu t}$ and $e^{(\Lambda-\mu)t}$. The exponents of these exponentials satisfy $\delta+k_2+\mu >\mu > -\Lambda+\mu$.

The variable $H$ tends to zero exponentially as $e^{-(\delta+k_2+\mu)t}$ faster than $R$ that tends as $e^{-\mu t}$. The evolution of $S$ depends on $\Lambda-\mu$.
\begin{itemize}
\item
If $\Lambda-\mu <0$, $S$ tends to zero as $e^{-|\Lambda-\mu|t}$.
In the presence of infection, it is expected that
$S$ tends also to zero.
\item
If $\Lambda-\mu >0$,  $S$ grows as $e^{|\Lambda-\mu|t}$. In the presence of infection, it is expected that $S$ grows also (probably more slowly) or it can even decrease if the infection is strong.
\item
If $\Lambda=\mu$, then $S$ tends to a constant. In the presence of infection, it is expected that it decreases at beginning and depends on the infection strength. This variable may tend to zero or a constant value.
\end{itemize}

Note that by the uniqueness of solution of system (\ref{my:eqf1}), if some of $E(0), I(0)$ or $A(0)$ are not zero there cannot be any time $t$ at witch these three variables vanish simultaneously.

The contagion-free manifold will be defined as the set

\begin{equation}
CFM=\{(S,0,0,0,H,R)^T, \hspace{8 pt} S,H,R\geq 0 \},
\end{equation}
and as we have observed, it is an invariant manifold.

Let us consider a point $Q^*$ in the $CFM$. The Jacobian matrix of the vector field at $Q^*$ can be written as follows

{\small
\begin{equation}
\notag
J=\begin{pmatrix}
\Lambda-\mu&0&-\frac{\beta S^*}{S^*+H^*+R^*}&-\frac{\beta S^*}{S^*+H^*+R^*}&0&0\\
0&-(\sigma +\mu)&\frac{\beta S^*}{S^*+H^*+R^*}&\frac{\beta S^*}{S^*+H^*+R^*}&0&0\\
0&p\sigma&-(\gamma+\mu)&0&0&0\\
0&(1-p)\sigma&0& -(k_1+\mu)&0&0\\
0&0&\gamma&0&-(k_2+\delta+\mu)&0\\
0&0&0&k_1&k_2&-\mu\\
\end{pmatrix}.
\end{equation}
}
This matrix has the eigenvalue $(\Lambda-\mu)$,  and its associated eigenvalue is $(1, 0, … , 0)^T$. This means that the component of the fundamental solution associated with the eigenvalue $(\Lambda-\mu)$ only affects the first variable $S$. The behaviour of the other variables will depend on the remaining five eigenvalues. The eigenvalue $-\mu$ has the eigenvector $(0, … , 0, 1)^T$ . The component associated with the
eigenvalue $-\mu$ only affects the variable $R$. Another eigenvalue is $-(k_2+\delta+\mu)$, whose eigenvector is $\left(0, 0, 0, 0, -\frac{k_2+\delta}{k_2}, 1 \right)$. The component of the solution associated with this eigenvalue only affects the variables $H$ and $R$. This means that the behaviour of the variables $E$, $I$ and $A$ around the point $Q^*$ depends only on the three remaining eigenvalues.

The remaining three eigenvalues are those of the main submatrix of $J$ of dimension 3 obtained by eliminating the first and the last two rows and columns. The trace of this submatrix is clearly negative and a necessary condition for the equilibrium point not to be unstable is that its determinant is negative, that is,

\begin{equation}
\begin{array}{l}

\beta\left[\cfrac{S^*}{S^*+H^*+R^*}\right]\left(\sigma(1-p)(\gamma+\mu)+p\sigma(k_1+\mu) \right)\\[8pt]-(k_1+\mu)(\gamma+\mu)(\mu+\sigma)<0,
\end{array}
\end{equation}
that can be written in the form

\begin{equation}\label{montifin11}
\cfrac{S^*}{S^*+H^*+R^*} R_0<1,
\end{equation}

with

\begin{equation} \label{montifin}
 R_0=\cfrac{\beta(k_1 p\sigma+ p \mu \sigma +(1-p) (\gamma +\mu )\sigma)}{(\gamma +\mu ) (k_1+\mu )(\mu +\sigma )}.
\end{equation}

Applying the Routh-Hurwitz criterion to the submatrix of $J$ associated to the variables $E, I, A$, we obtain that the eigenvalues of this submatrix have negative real part if and only if \eqref{montifin11} is satisfied. If we impose the condition to be satisfied for all points $Q^*$ in the $CFM$, then it must be $R_0 <1$, that is, the $CFM$ is locally stable if $R_0<1$.

If we apply the new generation matrix method reported in \citep{diekmann2010, perasso2018} at a point $Q^*$ in the $CFM$. we obtain the matrices

\[
\begin{array}{l}
{\cal F}=\begin{pmatrix}
 0 & \frac{\beta S^*}{S^*+H^*+R^*} & \frac{\beta S^*}{S^*+H^*+R^*} \\
 0 & 0 & 0\\
 0 & 0 & 0 
\end{pmatrix}, \\[8pt]
{\cal V}^{-1}=\begin{pmatrix}
 \cfrac{1}{(\mu+\sigma)} & 0 & 0\\
 \cfrac{\sigma p(k_1+\mu)}{(k_1+\mu)(\gamma+\mu)(\mu+\sigma)} & \cfrac{1}{(\gamma+\mu)} & 0\\
 -\cfrac{(p-1)\sigma}{(k_1+\mu)(\mu+\sigma)} & 0 & \cfrac{1}{(k_1+\mu)}
\end{pmatrix}
\end{array},
\]

The basic reproduction number is given by the maximum absolute value of the eigenvalues of the matrix ${\cal F} {\cal V}^{-1}$, that is, its spectral radious. In this case we obtain the expression in \eqref{montifin11} and if we consider the maximum value in $S^*$,$H^*$,$R^*$,  that is in the whole $CFM$, we obtain $R_0$ in \eqref{montifin} as the basic reproductive number.

Let's analyze when this variety is globally asymptotically stable, that is, under what conditions on the parameters $E(t) \to 0 $, $I(t) \to 0 $, $A(t) \to 0 $ regardless of the values taken by $S$, $H$, $R$. Consider the equations associated with $E$, $I$, $A$

\begin{equation}
\begin{cases}
\vspace{2mm}
\cfrac{dE}{dt}=\cfrac{\beta S(I+A)}{N}-(\sigma+\mu)E\\
\vspace{2mm}
\cfrac{dI}{dt}=p\sigma E -(\gamma+\mu)I\\
\vspace{2mm}
\cfrac{dA}{dt}=(1-p)\sigma E -(k_1+\mu)A
\end{cases}
\end{equation} 

These three equations can be rewritten as

\begin{equation}\label{lastmodifi}
y'=My-v.
\end{equation}

where

\begin{equation}
y=(E, I, A)^T
\end{equation}

and
{ \small
\begin{equation}\label{montijanomodifi2}
M=\begin{pmatrix}
-(\sigma+ \mu) & \beta & \beta\\
p \sigma & -(\gamma+\mu) & 0 \\
(1-p)\sigma & 0 & -(k_1+\mu)
\end{pmatrix},
\hspace{2 pt}
v=\begin{pmatrix}
g(t)\\
0\\
0
\end{pmatrix},
\end{equation}
}
with

\begin{equation}
\notag
g(t)=\beta (I(t)+A(t))\left(1-\cfrac{S(t)}{N(t)} \right).
\end{equation}
It is clear that $g(t)\geq 0 \hspace{8 pt} \forall t$.

In order to prove the global stability,  we will make use of the following comparison Theorem

\begin{teo}Let us suppose that \label{montijanorevi2}
\begin{itemize}
\item
 $x=(x_1 \ldots,x_n)^T$ is a vector of $R^n$.
\item
 $f(t,x)$ is a Lipschitz continuous function in $[t_0,\infty) \times {\cal E}$, ${\cal E}$ open subset of $R^n$, which is quasi monotonic, non decreasing, that is, each component of $f_i(t, x_1, \ldots, x_n)$ is monotonic non decreasing in the variables $x_j$ with $j \ne i$.
\item
 $y(t$) is a differentiable function in $R$  with $y(t) \in {\cal E}$.
\item
 $z(t)$ is a solution of $z'=f(t,z)$.
\item
 $y_i(t_0) \leq z_i(t_0)$\hspace{3mm} $\forall i$.
\item
 $y_i'(t)\leq f_i(t,y)$ \hspace{3mm} $\forall i$\hspace{3mm} and \hspace{3mm} $\forall y\in {\cal E}$. 
\end{itemize}
Then $y_i(t)\leq z_i(t), \hspace{3mm} i=1,2,\ldots, n \hspace{3mm} \forall t $.
\end{teo}
\Proof
This result is a particular case of Corollary 1.7.1 in  \citep{Lakshmikantham1969}  and also of comparison Theorems in \cite{McNabb1986}.
\Endproof

Then, we can state the following stability result

\begin{teo}\label{finteo}
If $R_0<1$, then $E(t) \to 0$, $I(t) \to 0$ and $A(t) \to 0$ and disease free manifold is globally stable.
\end{teo}
\begin{Proof}
Let us apply Theorem \ref{montijanorevi2} with $y(t)$ the solution of \eqref{lastmodifi} and $f(t,y)=My$. It is clear that $My$ is a cuasi monotonic non decreasing function since the non diagonal elements of $M$ are non negative. It is also clear that $y_i' \leq f_i(t,y)$ because $g(t)$ is greater or equal to zero. Taking $y(0)=z(0)$ we conclude that $y_i(t) \leq z_i(t)$ for all $t>0$. Now let us note that the matrix $M$ is in fact the submatrix of $J$ obtained by eliminating the first and the last two rows and columns and putting  $H^*=R^*=0$. Then if $R_0<1$, the three eigenvalues of $M$ have real part smaller than $0$ and $z(t)\to 0$. Consequently, $y(t)\to 0$ (note that $y_i(t)\geq 0$). This proves the global stability of the contagious free manifold. Additionally, if we include in Eq. \eqref{lastmodifi} the equation (linear) of system \eqref{my:eqf1} corresponding to $H(t)$, with the same reasoning we obtain that $H(t)\to 0$, and the disease free contagious manifold is globally asymptotically stable.
\end{Proof}

\subsection{Stability of the equilibrium points}

\subsubsection{Case $\Lambda > \mu$}

If parameters of the system \eqref{my:eqf1} are all positive, this system of equations has only one critical point $ P^*=(0, 0, 0, 0, 0, 0)^T$ . Nevertheless, the Jacobian matrix of the system has some components with the term

\begin{equation}
\notag
\cfrac{\beta S (S+E+A+H+R)}{\left(S+E+I+A+H+R\right)^2}
\end{equation}
which is not continuous when variables tend to zero. Therefore, we cannot use linearization to study the local stability of the origin. 

We will obtain next some conditions on the parameters of the model and the initial conditions for which the total population grows with time and therefore the origin can not be stable.

\begin{teo}\label{teorev1}
If $\Lambda>\mu$ and $R_0<1$, there exist a constant $K$ such that if $N(0)>K$, then the total population $N(t)$ tends to infinity when $t\to\infty$.
\end{teo}

\begin{Proof}
Let us consider the linear system

\begin{equation}\label{mont2}
\begin{cases}
\vspace{2mm}
\cfrac{d\hat{E}}{dt}=\beta \hat{I}+\beta \hat{A}-(\sigma+\mu)\hat{E}\\
\vspace{2mm}
\cfrac{d\hat{I}}{dt}=p\sigma \hat{E} -(\gamma+\mu)\hat{I}\\
\vspace{2mm}
\cfrac{d\hat{A}}{dt}=(1-p)\sigma \hat{E} -(k_1+\mu)\hat{A}\\
\vspace{2mm}
\cfrac{d\hat{H}}{dt}=\gamma \hat{I} -k_2 \hat{H}-(\delta+\mu)\hat{H}\\
\end{cases}.
\end{equation}
whose general solution has the form

\begin{equation}
\hat{P}(t)=C_1\vec{v_1}e^{\lambda_1 t}+C_2\vec{v_2}e^{\lambda_2 t}+C_3\vec{v_3}e^{\lambda_3 t}+C_4\vec{v_4}e^{\lambda_4 t},
\end{equation}
where $\lambda_1, \lambda_2, \lambda_3$ are the eigenvalues of the matrix $M$ in \eqref{montijanomodifi2}, $\lambda_4= -(k_2+\delta+\mu)$ and $v_i$ are the corresponding eigenvectors. Proceeding as in Theorem \ref{finteo}, the components $E, I, A, H$ of the solution of system \eqref{my:eqf1} are upper bounded by the ones of \eqref{mont2} and in particular

\begin{equation}
H(t)\leq \hat{H}(t)=B_1e^{\lambda_1 t}+B_2e^{\lambda_2 t}+B_3e^{\lambda_3 t}+B_4e^{\lambda_4 t}.
\end{equation}
where $B_i$ are constants that depend on the eigenvectors $v_i$ and the initial conditions.

Denoting $u = max_i {Re \lambda_i}$, if $R_0 <1$, then $u <0$ , $H(t)$ goes to zero and there exists a constant $B$ such that

\begin{equation}\label{mont3}
H(t)\leq Be^{ut}, \hspace{2 mm} u<0
\end{equation}
Adding the equations of system \eqref{my:eqf1}, we get

\begin{equation}\label{mont1}
\hbox{d}N/\hbox{d}t= (\Lambda  -\mu) N - \delta H.
\end{equation}
If the bound \eqref{mont3} is substituted in \eqref{mont1} we get

\begin{equation}
\cfrac{dN}{dt} \geq (\Lambda-\mu)N - B e^{u t}
\end{equation}
which implies

{\small
\begin{equation}
N(t)\geq \left[N(0)-\cfrac{\delta B}{(u-\Lambda+\mu)}\right]e^{(\Lambda-\mu)t}+\cfrac{\delta B}{(u-\Lambda+\mu)}e^{ut},
\end{equation}}
and since $\Lambda-\mu>0$, if  $N(0) >\frac{\delta B}{(u-\Lambda+\mu)}$, $N(t)$ tends to infinity.

Note that the constant $B$ depends on the initial conditions $E(0), I(0),$ $A(0), H(0)$ and therefore $S(0)+R(0)$ must be large enough with respect to the other initial conditions.

\end{Proof}

The above result let us know that the population grows with time if $R_0 <1$ and the initial population is large enough. We will see next that under a little more restrictive condition on $R_0$, the total population grows with time whatever the initial population is.

\begin{teo} If $\Lambda-\mu>0$ and\label{teorev2}
\begin{equation} \label{montirevinf1}
\hat R_0 =\dfrac{\beta \sigma[k_1 p + (1-p) \gamma]}{\gamma k_1 \sigma} <1,
\end{equation}
then $S(t)$ tends to infinity when $t \to \infty$.
\end{teo}

\begin{Proof} 
From \eqref{my:eqf1} we have

\begin{equation}
R'(t)=k_1A+K_2H-\mu R(t)\geq -\mu R(t),
\end{equation} 

There exist a point $t_0$ for which $R(t_0) >0$ (note that if $R(0)=0$, from Theorem \ref{teo1} $R(t) >0$ in a neighbourhood of $0$). Then we have the bound

\begin{equation}\label{montijanorev3}
R(t)\geq R(t_0)e^{-\mu t}, \hspace{2 mm}  \forall t\geq t_0
\end{equation}

Now, note that $R_0$ is a decreasing monotone function with respect to $\mu$. Therefore $R_0  \leq \hat R_0 <1$ for any $\mu >0$ and Theorem \ref{finteo} applies. There exist a constant $K$ such that

\begin{equation}\label{montijanomodificado33}
I(t)+A(t)\leq Ke^{ut} \hspace{33 mm} \forall t \geq t_0
\end{equation}
$u$ being the maximum real part of the eigenvalues of matrix $M$ in \eqref{montijanomodifi2}. Using the bounds \eqref{montijanorev3} and \eqref{montijanomodificado33} in the first equation of  \eqref{my:eqf1}, we obtain
{ \small
\begin{equation}\label{montijamomodifi7}
\begin{array}{l}
S'(t)\geq (\Lambda-\mu)S-S\beta\cfrac{I+A}{I+A+R}\\
S'(t)\geq S\left[\Lambda-\mu-\beta\cfrac{Ke^{ut}}{Ke^{ut}+R(t_0)e^{-\mu t}} \right] \hspace{2 mm} \forall t \geq t_0\\
\end{array}
\end{equation}}

Since $\hat R_0 <1$ then $u < -\mu$ and then

\begin{equation}\label{montijanomodifi71}
\cfrac{Ke^{ut}}{Ke^{ut}+R(t_0)e^{-\mu t}} \to 0.
\end{equation}

There will  exist a time $t^*$ and a positive constant C such that 

\begin{equation}
\Lambda-\mu-\beta\cfrac{Ke^{ut}}{Ke^{ut}+R(t_0)e^{-\mu t}} \geq C >0 \hspace{2 mm} \forall t \geq t^*.
\end{equation}

Therefore

\begin{equation}
S'(t)\geq C S(t) \hspace{33 mm} \forall t \geq t^*,
\end{equation}
which implies that $S(t) \to \infty$

Recall that $u= max Re \lambda_i$. where $\lambda_1, \lambda_2, \lambda_3$ are the eigenvalues of $M$ and $\lambda_4 = - (k_1+\delta+\mu) < \mu$. Then the condition $u<-\mu$ is equivalent to say that the real parts of the eigenvalues of $M$  are smaller than $-\mu$. But the eigenvalues of $M$ are precisely the eigenvalues of the matrix $M$ with $\mu=0$ minus $\mu$, that is

$\lambda_i(\mu) = \lambda_i(\mu=0)-\mu$,\\
which implies that $u<-\mu$ if $Re\lambda_i(mu=0) <0$.

The eigenvalues $\lambda_i(\mu=0)$ have real part smaller than $0$ if and only if $\hat{R}_0 = R_0(\mu=0) <1$ and this proves the theorem.
\end{Proof}

\subsubsection{Case $\Lambda < \mu$}
First, by adding the equations of system \eqref{my:eqf1}, we get
\[
\hbox{d}N/\hbox{d}t= (\Lambda  -\mu) N - \delta H.
\]

If parameters of the system \eqref{my:eqf1} are all positive, this system of equations has only one critical point $ P ^ * = (0,0,0,0,0,0) ^ T $. The Jacobian matrix of the system is not continuous at zero and we can not use linearization to study the local stability. Nevertheless it is possible to see that this point is globally stable.

\begin{teo} (Global stability of $P^*$) \label{teomontifin1}
If at $t=0$, all variables are non negative and $\Lambda < \mu$, they tend to zero when $t\to\infty$.
\end{teo}
\Proof
As $\Lambda < \mu$, if $H\ge 0$, the total population $N$ decreases and tends to zero as $t$ tends to infinity. By Theorem \ref{teo1}, $H(t)\ge 0$, which implies that $N'(t) \leq (\Lambda-\mu)N(t)$. Therefore, $N(t)\to 0$ when $t\to\infty$, and since all variables are non negative, they must tend to zero.
\Endproof

This result confirms that the proposed model is consistent with the biological system, because all variables in system \eqref{my:eqf1} tend to zero, but does not provide information on the epidemic behaviour. In practice, the values of $\Lambda$ and $\mu $ are small. If $ \Lambda < \mu $, total population will tend to die after a long time, but what we want to know is if the epidemic remits ($E, I, A, H$ tend to zero) in a reasonable period of time, leaving only the $ S $ and $ R $ populations, or the epidemic becomes endemic ($ E, I, A, H $ tend to zero slowly, at a speed similar to that of $ S $), or if the epidemic affects the entire population ($ S, E, I, A, H $ tend to zero in a reasonable period of time). According to theorem \ref{finteo} if the basic reproduction number $R_0$ is smaller than 1, the infection remits exponentially at least as $e^{u t}$, $u$ being the largest real part of matrix $M$ in \eqref{lastmodifi}. If $u < -\mu$, the infection remits faster than the natural dead speed, that is, the disease remits (see \eqref{montijamomodifi7} and use \eqref{montijanomodifi71}). However, if $u$ is close to $-\mu$, the infection can stay active for a long time. Additional information can be obtained by studying the cases $ \Lambda = \mu $ and $ \Lambda = \mu = 0 $.

Note that if we compute $R_0$ by applying the new generation matrix methods reported in \citep{diekmann2010, perasso2018} at the null equilibrium point we obtain again the value \eqref{montifin}. 

\subsubsection{Case $\Lambda = \mu$}
In this case, the total population satisfies
\[
\hbox{d}N/\hbox{d}t=  - \delta H,
\]
which implies that $N(t)$ is non negative and decreases if $H\geq 0$. This means that $N(t)$ has a limit $N^*\ge 0$ when $t\to\infty$.

If parameters of the system \eqref{my:eqf1} are positive, any point $P^*=(S^*,0,0,$ $0,0,0)^T$ is an equilibrium point. The Jacobian matrix of the system at the equilibrium point (limit of the Jacobian matrix at a point $ P $ when it tends to $ P^* $)  with $S^*>0$ can be written can be written as follows

\[
J=\begin{pmatrix}
0 & \mu & \mu-\beta& \mu-\beta & \mu & \mu \\
0 & -(\sigma+\mu) & \beta & \beta & 0 & 0 \\
0 & p\sigma & -(\gamma+\mu) & 0 & 0 & 0 \\
0 & (1-p)\sigma & 0 & -(k_1+\mu) & 0 & 0 \\
0 & 0 & \gamma & 0 & -(k_2+ \delta+\mu) & 0 \\
0 & 0 & 0 & k_1 & k_2 & -\mu \\
\end{pmatrix}.
\]

This matrix has the eigenvalue $0$, the eigenvector associated with this eigenvalue is $ (1, 0, \ldots, 0)^T $. This means that the component of the fundamental solution associated with the eigenvalue 0 only affects the first variable $ S $. The behaviour of other variables will depend on the remaining five eigenvalues. The eigenvalue $ - \mu $ has the eigenvector $ (- 1, \ldots, 0,1)^T $. The component associated with the eigenvalue $ - \mu $ only affects the variables $ S $ and $ R $. Another eigenvalue is $ - (\delta + k_2 + \mu) $, whose eigenvector is $ (- \frac {\delta \mu} {k_2 (\delta + k_2 + \mu)}, 0, 0, 0, -\frac {k_2+\delta} {k_2 },1)^T $. The component of the solution associated with this eigenvalue only affects the variables $ S $, $ H $ and $ R $. This means that the behaviour of the variables $ E, I$ and $A$ depends on the three remaining eigenvalues.

The remaining three eigenvalues are those of the main submatrix of dimension 3 obtained by eliminating the first and the last two rows and columns. But this submatrix is the same as the one studied in section \ref{lastref}, but with $H^*=R^*=0$. There we concluded that their three eigenvalues have negative real part $(u<0)$ if and only if $R_0 <1$, with $R_0$ given by equation \eqref{montifin}. This value coincides with the one obtained by the new generation matrix method (see \cite{diekmann2010, perasso2018}).

If $R_0 >1$, all the equilibrium points (except possibly the zero point)  are unstable.  If $R_0 <1$, since the Jacobian matrix has an eigenvalue zero, we can not conclude their stability.   

Note that at the origin the Jacobian matrix of the system (\ref{my:eqf1}) is not defined (there is no limit when all variables tend to zero) and the linearization theorem cannot be applied. The following results give some insight about their stability.

\begin{lema}\label{lema1}
Let $ f(t) $ be a twice differentiable function, with a second derivative bounded at $[0,\infty)$. If $\lim_{t\to\infty}| f(t)|=f^*<\infty$ then $ \lim_{t\to\infty} f'(t)=0 $.
 \end{lema}
\Proof
First, if $f'(t)$ has a limit $f'^*(t)\neq 0$ when $t \to \infty$, we can ensure that for large $t$ $\left| f'(t) \right|>M$ that implies that $f(t)$ cannot be bounded. If $f'(t)$ has no limit, the upper and lower limit are real numbers, not equal to each other and there exist a constant $\alpha$ such that,
\begin{equation}
\notag  \liminf_{t \to \infty} f'(t)< \alpha < \limsup_{t \to \infty} f'(t)
\end{equation}

In addition,  $\liminf f'(t)\cdot \limsup f'(t)\leq 0$ because otherwise we ensure that $\left| f'(t) \right|>M$ for large enough $t$.  From the definition of upper limit,  there exists a sequence of values of $f(t)$ approaching the upper limit,  and another sequence of values of $f(t)$ approaching the lower limit.  Let us suppose that $\alpha >0$ (the case $\alpha<0$ is similar).  There exist infinite sequences $\{t_k\}$ and $\{s_k\}$,  such that,

\begin{equation}
\begin{cases}
\notag
k=1,2,3,4..... \\
t_1 < s_1 < t_2 < s_2 <...\to \infty \\
f'(t_k) > \alpha \\
f'(s_k) = \cfrac{\alpha}{2} \\
f'(t) \geq \cfrac{\alpha}{2} \hspace{3 mm} t \in[t_k,s_k].\\
\end{cases}
\end{equation}

The second derivative is bounded by a positive constant $K$ and applying the mean value theorem, we obtain 
\begin{equation}
\notag  \cfrac{\alpha}{2}\leq\left|f'(t_k)-f'(s_k)\right| \leq K \left|t_k-s_k \right|,
\end{equation}
therefore,
\begin{equation}
\notag  \left|t_k-s_k\right| \geq \cfrac{\alpha}{2K}.
\end{equation}

Applying again the mean value theorem, for some $c\in (t_k, s_k)$

\begin{equation}
\notag \left|f(t_k)-f(s_k)\right| =f'(c)\left|t_k-s_k \right|\geq \cfrac{\alpha^2}{4K}.
\end{equation}

But this inequality contradicts the Cauchy condition, which states that

\begin{equation} 
 \begin{array}{l}
\notag \lim_{t \to \infty} \left| f(t)-f(s) \right|=0, \\
\hspace{3.2 mm}\tiny{s>t}
 \end{array}
\end{equation}

and consequently,

\begin{equation}
\notag \lim_{t\to\infty} inf f'(t)=0, \hspace{3 mm} \lim_{t\to\infty} sup f'(t)=0.
\end{equation}

\Endproof

 \begin{lema} \label{lemma2}
 If the initial conditions are non negative, $P(0)\in \mathcal{D}$, and $\Lambda=\mu$, then $S(t)$, $E(t)$, $I(t)$, $A(t)$, $H(t), R(t)$ are bounded and have two bounded derivatives in $[t_0,\infty]$.
 \end{lema}
\Proof
It is a direct consequence of the form of system (\ref{my:eqf1}), the boundedness of $N$ and theorem \ref{finteo}.
\Endproof
 \begin{lema} \label{lemma3}
 If the initial conditions are non negative, $P(0)\in \mathcal{D}$, and $\Lambda=\mu$, then $N(t)$ has a limit $N^*\ge 0$ when $t\to\infty$ and  $\lim_{t\to\infty}N'(t)=0$.
 \end{lema}
\Proof
It is consequence of Lemma \ref{lema1}.
\Endproof

 \begin{lema} \label{lemma4}
 If $\Lambda=\mu\neq 0$ and $\lim_{t\to\infty} S(t)=0$, then $\lim_{t\to\infty}N(t)=0$ and all variables tend to zero  when $t$ tends to infinity.
 \end{lema}
 \Proof
 From Lemma \ref{lema1}, if $S(t)$ tends to zero, then  $S'(t)$ tends to zero, and using the first equation in \eqref{my:eqf1},  $N(t)$ tends to zero.
 \Endproof
 
\begin{teo} \label{otrarevi1}
Let us consider the system \eqref{my:eqf1} with initial condition $P(0)\in \mathcal{D}$, and $\Lambda=\mu\neq 0$. Then
\begin{itemize}
\item
Any solution tends to an equilibrium point $(S^*,0,0,0,0,0)$ when $t\to\infty$.
\item
 If \eqref{montirevinf1} is satisfied, then any solution whose initial condition does not vanish tends to an equilibrium point with $S^*>0$.  The origin is unstable.
 \item
 If $R_0>1$, then  $N(t)\to 0$ when $t \to \infty$.  The origin (equilibrium point with $S^*=0$) is globally asymptotically stable.
 \item
 If $R_0<1$ then the equilibrium points $P^*$ with $S^*>0$ are stable.
\end{itemize}
\end{teo}

\Proof
According to lemma \ref{lemma3}, $ N(t)$ has limit $N^* \ge 0$ and its derivative tends to zero when $ t $ tends to infinity. Therefore, $ H(t)$ also tends to zero, and since its second derivative is bounded at $ [t_0, \infty]$, we can ensure that $ H'(t) \to 0$. This implies that $ I(t) $ tends to zero and therefore also $I'(t) \to 0 $. From the third of the differential equations, $ E(t) $ should tend to zero, so that $ E'(t) \to 0 $, and hence either $ S(t) \to 0 $, which implies (by lemma \ref{lemma4}) that the solution tends to the origin, or $ A(t) \to 0 $. In this case, $ E(t), I(t), A(t), H(t) \to 0 $, so $ R(t) \to 0 $ and this implies that the solution tends to an equilibrium point $P^*$.

From the first equation of system \eqref{my:eqf1} it is clear that

\begin{equation}
S' \geq \Lambda R - \beta (I+A)
\end{equation}

Proceeding as in Theorem \ref{teorev2} and using the bounds \eqref{montijanorev3} \eqref{montijanomodificado33}, we obtain

\begin{equation}
S' \geq R(t_0)e^{-\mu t}\left[\Lambda-\beta \cfrac{ke^{ut}}{R(t_0)e^{-\mu t}} \right].
\end{equation}
where $u$ is again the maximum real part of the eigenvalues of the matrix $M$ in \eqref{montijanomodifi2}.

If \eqref{montirevinf1} is satisfied, then $|u| > |\mu|$ and

 \begin{equation}
  \cfrac{ke^{ut}}{R(t_0)e^{-\mu t}} \to 0
 \end{equation}
 There will exist a time $t^*$ such that
 \begin{equation}
  \left[\Lambda-\beta \cfrac{ke^{ut}}{R(t_0)e^{-\mu t}} \right] \geq D > 0\  \forall t \geq t^*,
  \end{equation}
  
Therefore, for $t>t^*$, $S(t)$ is an increasing and upper bounded function that must have a limit $S^*>0$. If $R_0 >1$, then all the equilibrium points with $S^*>0$ are unstable.

To prove the stability of $P^*$, let us see that for $\delta>0$ there exists $\epsilon>0$ such that if  $\parallel P(0)-P^* \parallel<\epsilon$ then $\parallel P(t)-P^* \parallel<\delta$ $\forall t$. Let us consider the maximum norm and $\parallel P(0)-P^* \parallel<\epsilon$. We must see that $\parallel S(t)-S^* \parallel<\delta$, and that for any other variable $X$, $|X(t)|<\delta$. Comparing the solution of \eqref{my:eqf1} with the one of the linear system \eqref{mont2}, as in theorem \ref{teorev2}, we obtain that
{\small \begin{equation}
X(t)\leq k e^{-\min\{|u|,\mu\}t} \hspace{1 mm}  \max\{E(0), I(0), A(0), H(0), R(0)\}
\end{equation} }
Since $\parallel P(0)-P^* \parallel<\epsilon$, we have that
\begin{equation}
X(t)\leq k \epsilon e^{-\min\{|u|,\mu\}t}\leq k\epsilon
\end{equation}
and in particular
\begin{equation}
I(t)+A(t)\leq 2k\epsilon e^{-\min\{|u|,\mu\}t} 
\end{equation}
Now, from the first equation of \eqref{my:eqf1}
\begin{equation}
-\beta(I(t)+A(t))\leq S'(t)\leq \Lambda(I(t)+A(t)), 
\end{equation}
which implies that
\begin{equation}
|S'(t)|\leq \max\{\beta,\Lambda\}2k\epsilon e^{-\min\{|u|,\mu\}t},
\end{equation}
Using the fundamental theorem of calculus 
\begin{equation}
\begin{array}{l}
\displaystyle |S(t)-S(0)|=\left|\int_0^t S'(x) dx\right| \leq \displaystyle \max \{\beta,\Lambda \}2k \epsilon \int_0^t e^{-\min\{|u|,\mu\}x}dx= \\ [2mm]
\displaystyle \max \{\beta,\Lambda \}k \epsilon \frac{1}{\min\{|u|,\mu\}}\left(1-e^{-\min\{\mu\}t}\right)\leq 
\displaystyle 2k\frac{\max \{\beta,\Lambda \}}{\min\{|u|,\mu\}}\epsilon=L\epsilon
\end{array}
\end{equation}
Taking \hspace{2mm}$\displaystyle \epsilon<\frac{\delta}{L+1}$ \hspace{2mm} and \hspace{2mm} $\displaystyle \epsilon< \frac{\delta}{k}$, we deduce

\begin{equation}
\begin{array}{l}
\left|S(t)-S^*\right|\leq\left|S(t)-S(0)\right|+\left|S(0)-S^*\right| \\ [2mm]
\hspace{18.5mm}\leq (L+1)\epsilon<\delta
\end{array}
\end{equation}  
And
\begin{equation}
\displaystyle  \left|x(t)\right|<\delta
\end{equation}
which proves the stability.
\Endproof

The above theorem proves that if $R_0>1$ all the equilibrium points with $S^*>0$ are unstable, but the origin is a globally stable point. The COVID-19 disease has a high transmission (due to SARS-CoV-2 virus is very contagious) and $S$ goes to zero quickly or well the infection remains active for a long time.

If $R_0 <1$, the theorem states the stability of the equilibrium poins with $S^*>0$, but does not states the stability or the unstability of the origin. However, if \eqref{montirevinf1} is satisfied, the origin is unstable. The infection disapears with time and the population tends to a stable value $N^* = S^*$.

\subsubsection{Case $\Lambda = \mu=0$}

This is a particular case of $\Lambda = \mu$ so the total population obeys the following equation 
\[
\hbox{d}N/\hbox{d}t=  - \delta H.
\]
Total population $N$ decreases if $H\ge 0$, this means that $N(t)$ has a limit $N^*\ge 0$. The difference with the general case $\Lambda=\mu$ is that in this case the critical points are  $P^*=(S^*,0,0,0,0,R^*)^T$ at which the Jacobian matrix of the system (\ref{my:eqf1}) is 
\[
J=\begin{pmatrix}
0 & 0 & -\frac{\beta S^*}{S^*+R^*}& -\frac{\beta S^*}{S^*+R^*} & 0 & 0 \\[7pt]
0 & -(\sigma) & \frac{\beta S^*}{S^*+R^*} & \frac{\beta S^*}{S^*+R^*} & 0 & 0 \\
0 & p\sigma & -(\gamma) & 0 & 0 & 0 \\
0 & (1-p)\sigma & 0 & -(k_1) & 0 & 0 \\
0 & 0 & \gamma & 0 & -(k_2+ \delta) & 0 \\
0 & 0 & 0 & k_1 & k_2 & 0 \\
\end{pmatrix}.
\]

This matrix has the eigenvalue 0 double. The eigenvectors associated with this eigenvalue are $(0, 0, \ldots, 1)^T$ and $(1, \ldots, 0, 0)^T$. This means that the component of the fundamental solution associated with the eigenvalue 0 only affects the first variable $S$ and the last variable $R$. The behaviour of other variables will depend on the remaining four eigenvalues.

Another eigenvalue is $-(\delta+k_2)$, whose eigenvector is $(0, 0, 0, 0, -\frac{k_2+\delta}{k_2},1)^T$. This means that the behaviour of the variables $E, I$ and $A$ depends on the three remaining eigenvalues, which are those of the main submatrix of dimension 3 obtained by eliminating the first and the last two rows and columns. This submatrix was already studied in section \ref{lastref}, where we concluded that their three eigenvalues have negative real part if and only if

\begin{equation}\label{montijanomodificado34}
\hat{R}_0 \dfrac{S^*}{S^*+R^*} <1,
\end{equation} 
with $\hat{R}_0$ the basic reproductive number with $\mu=0$, that is

\begin{equation}\label{montijanomodificado35}
\hat{R}_0=R_0(\mu=0)=\dfrac{\beta[\gamma  (1 - p)+ k_1 p]  }{k_1  \gamma},
\end{equation}
which means that the equilibrium point is stable if 
 
\begin{equation} \label{ultiarre}
 \dfrac{S^*+R^*}{S^*}>\hat{R}_0.
\end{equation}

 The global stability of the equilibrium points is studied in a similar way to the case $\Lambda=\mu$, so the following theorem can be stated. 
 \begin{teo} \label{montifininfteo}
All solution of the differential system (\ref{my:eqf1}) with $\Lambda=\mu=0$, whose initial condition has all its components non negative, tends to an equilibrium point $(S^*, 0, 0,0,0, R^*)^T$ when $t$ tends to infinity.
 \end{teo}

In this case, if we apply the method employed in section \ref{lastref} we obtain precisely the condition \eqref{montijanomodificado34} 

Taking into account the local stability result, given some initial condition, the limit point $S^*$ and $R^*$ will be such that Eq. (\ref{ultiarre}) is satisfied. In other words, the solution will tend to an equilibrium point such that the limit point $P^*$ will be such that $S^*$ and $R^*$ satisfy Eq. (\ref{ultiarre}). As $R_0$ increases, $S^*$ decreases and the Eq. (\ref{ultiarre}) is fulfilled. 

\subsection{Statistical} \label{statt}
We use the simulation annealing algorithm for model fitting \cite{ingber1993}, in this process the sum of the squared errors between the original data and predicted values by model are minimized to obtain the best parameters for this fit. The quality of model fitting is verified by criteria for model assessment: $r^2$, $SSE$, $RMSE$ \cite{Gonzalez2017} and relative error (RE) \cite{peng2020},
\begin{equation}\label{SSE}
SSE=\sum_{j=1}^n\left(x_j-x_{jref}\right)^2,
\end{equation}

\begin{equation}\label{r}
1-r^2=\cfrac{\sum_{j=1}^n\left(x_j-x_{jref}\right)^2}{\sum_{j=1}^n\left(x_{jref}\right)^2-\cfrac{1}{n}\sum_{j=1}^n\left(x_{jref}\right)^2},
\end{equation}

\begin{equation}\label{rmse}
RMSE=\sqrt{\cfrac{\sum_{j=1}^{n}\left(x_j-x_{jref}\right)^2}{n}},
\end{equation}

\begin{equation}\label{RE}
RE=\sqrt{\cfrac{\left||x-x_{ref}|\right|_2}{\left||x_{ref}|\right|_2}}.
\end{equation}
\newpage
In Eq. (\ref{SSE})-(\ref{RE}), $x_{ref}$ are data daily provided by the Ministry of Health, whereas $x$ are those obtained from the model fitting, and $n$ is the number of experimental data.

\subsection{Simulation}
In this section, we will validate theoretical results obtained in section 2. We will consider the case $\Lambda>\mu$ with $\Lambda=0.015$ and $\mu=0.01$, the case $\Lambda<\mu$ with $\Lambda=0.008$ and $\mu=0.01$, the case $\Lambda=\mu\neq 0$ with  $\Lambda=0.008$, and the case $\Lambda=\mu=0$. Three values of $\beta$ are chosen for each one of these cases in such a way that the corresponding values of $R_0$ (Eq. \ref{montifin} and \ref{montijanomodificado35}) are smaller than, approximate equal to or larger than 1 (see Table \ref{my:tabla1}).

\begin{table}[h]
\centering
\caption{Values of basic reproduction numbers $R_0$ for simulation}
\label{my:tabla1}
\begin{tabular}{c| c c c c }
\hline
\multirow{2}{*}{$\beta$ (days$^{-1}$)} & $\Lambda>\mu$ &$\Lambda<\mu$ & $\Lambda=\mu$& $\Lambda=\mu=0$\\
\cline{2-5}
&$R_0$&$R_0$&$R_0$&$R_0$\\
\hline
0.05&0.3391&0.3391&0.3496&0.3975\\
0.20&1.3565&1.3565&1.3983&1.5910\\
1.00&6.7824&6.7824&6.9915&7.9500\\
\hline
\end{tabular}

\end{table}

The other parameters of the model and the initial conditions for those exposed, infected, infected asymptomatic, hospitalized and recovered cases are kept constant (Table \ref{my:tabla2}).

\begin{table}[h]
\centering
\caption{Value of each parameter for simulation and calculation of  $R_0$}
\label{my:tabla2}
\begin{tabular}{l c}
\hline
Parameters & Values\\
\hline
Population & 11380\\
$S(t=0)=S(0)$ & 10000\\
$E(t=0)=E(0)$ & 1000\\
$I(t=0)=I(0)$ & 100\\
$A(t=0)=A(0)$ & 80\\
$H(t=0)=H(0)$ & 200\\
$R(t=0)=R(0)$ & 0\\
$p$ & 0.5500\\
$k^{-1}_1$ (days) & 14\\
$\sigma^{-1}$ (days) & 5.1000\\
$k^{-1}_2$ (days) & 20\\
$\delta^{-1}$ (days) & 15\\
$\gamma^{-1}$ (days) & 3 \\
\hline
\end{tabular}
\end{table}

Parameters values in table 2 are taken arbitrarily, except $\sigma^{-1}$ (in days) and $ k^{-1}_1$ (in days) that are those reported in  \cite{lauer2020}. Parameter values are kept constant to perform the calculation of $R_0$ and simulations of temporal behaviour of symptomatic (blue line), asymptomatic (dashed green line), hospitalized patients (red line), susceptible (magenta line) and recovered (black line) cases, as shown in Fig. \ref{my:figf2} to Fig. \ref{my:figf9}. For numerical solution of the differential system \eqref{my:eqf1}, the ODE45 algorithm implemented in the Matrix Laboratory (MATLAB) is used \cite{VRABEL2015}.

\subsubsection{Case $\Lambda>\mu$}
\begin{figure}[ht]
\centering
\includegraphics[scale=2.7]{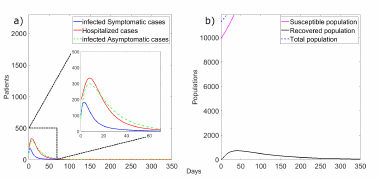}
\caption{\small Simulation of system (\ref{my:eqf1}) for $\Lambda=0.015$, $\mu=0.01$, $ \beta=0.05\ days^{-1}$ and $R_0=0.3391$. a) Infected symptomatic (blue line), hospitalized (red line) and infected asymptomatic (dashed green line) cases b) Susceptible (magenta line), recovered (black line) and total populations (dashed blue line).}
\label{my:figresp1}
\end{figure}

\begin{figure}[ht]
\centering
\includegraphics[scale=2.7]{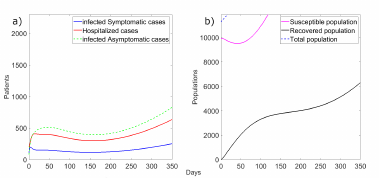}
\caption{\small Simulation of system (\ref{my:eqf1}) for $\Lambda=0.015$, $\mu=0.01$, $ \beta=0.2\ days^{-1}$ and $R_0=1.3565$. a) Infected symptomatic (blue line), hospitalized (red line) and infected asymptomatic (dashed green line) cases b) Susceptible (magenta line), recovered (black line) and total populations (dashed blue line).}
\label{my:figresp2}
\end{figure}

\begin{figure}[ht]
\centering
\includegraphics[scale=2.7]{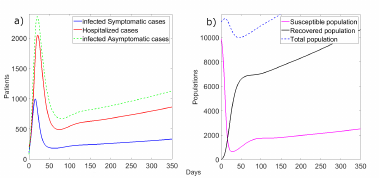}
\caption{\small Simulation of system (\ref{my:eqf1}) for $\Lambda=0.015$, $\mu=0.01$, $ \beta=1\ days^{-1}$ and $R_0=6.7824$. a) Infected symptomatic (blue line), hospitalized (red line) and infected asymptomatic (dashed green line) cases b) Susceptible (magenta line), recovered (black line) and total populations (dashed blue line).}
\label{my:figresp3}
\end{figure}

Fig. \ref{my:figresp1}a, Fig. \ref{my:figresp2}a and Fig. \ref{my:figresp3}a show the temporal behaviour of infected symptomatic  (blue line), hospitalized (red line) and infected asymptomatic (dashed green line) cases while in Fig. \ref{my:figresp1}b, Fig. \ref{my:figresp2}b and Fig. \ref{my:figresp3}b the susceptible (magenta line), recovered (black line), and the total populations (dashed blue line) are shown.

\newpage

The simulations reveal that the total population and the number of susceptible people tend to infinity, and the infected, asymptomatic and hospitalized cases tend to zero when $R_0 <1$, in agreement with Theorem \ref{teorev1}, Theorem \ref{teorev2} and Theorem \ref{finteo} respectively. When $R_0 >1$, the susceptible tends to infinity and the infected, asymptomatic and hospitalized  grow with time. The number of infected people increases and the growth velocity of susceptible people decreases faster as $R_0$ increases. The number of hospitalized cases depend on the values of $\gamma$, $\beta$ and $R_0$, showing different scenarios. First, hospitalized cases depend on the symptomatic and asymptomatic cases as show in Fig. \ref{my:figresp1}a, Fig. \ref{my:figresp2}a and Fig. \ref{my:figresp3}a. Second, the number of hospitalized cases is lower than that the asymptomatic cases when $R_0 \geq 1$, as shown in Fig. \ref{my:figresp2}a and \ref{my:figresp3}a.

\subsubsection{Case $\Lambda<\mu$}
\begin{figure}[h]
\centering
\includegraphics[scale=2.7]{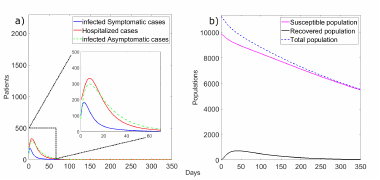}
\caption{\small Simulation of system (\ref{my:eqf1}) for $\Lambda=0.008$, $\mu=0.01$, $ \beta=0.05\ days^{-1}$ and $R_0=0.3391$. a) Infected symptomatic (blue line), hospitalized (red line) and infected asymptomatic (dashed green line) cases b) Susceptible (magenta line), recovered (black line) and total populations (dashed blue line).}
\label{my:figf2}
\end{figure}

\begin{figure}[h]
\centering
\includegraphics[scale=2.7]{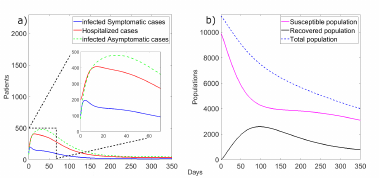}
\caption{\small Simulation of system (\ref{my:eqf1}) for $\Lambda=0.008$, $\mu=0.01$, $ \beta=0.2\ days^{-1}$ and $R_0=1.3565$. a) Infected symptomatic (blue line), hospitalized (red line) and infected asymptomatic (dashed green line) cases b) Susceptible (magenta line), recovered (black line) and total populations (dashed blue line).}
\label{my:figf3}
\end{figure}

\begin{figure}[h]
\centering
\includegraphics[scale=2.7]{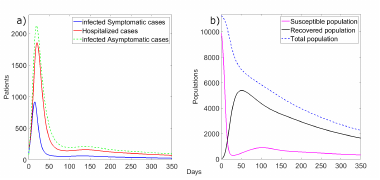}
\caption{\small Simulation of system (\ref{my:eqf1}) for $\Lambda=0.008$, $\mu=0.01$, $ \beta=1\ days^{-1}$ and $R_0=6.7824$. a) Infected symptomatic (blue line), hospitalized (red line) and infected asymptomatic (dashed green line) cases b) Susceptible (magenta line), recovered (black line) and total populations (dashed blue line).}
\label{my:figf4}
\end{figure}

Fig. \ref{my:figf2}a, Fig. \ref{my:figf3}a and Fig. \ref{my:figf4}a show the temporal behaviour of infected symptomatic  (blue line), hospitalized (red line) and infected asymptomatic (dashed green line) cases while in Fig. \ref{my:figf2}b, Fig. \ref{my:figf3}b and Fig. \ref{my:figf4}b the susceptible (magenta line), recovered (black line), and the total populations (dashed blue line) are shown.
\newpage

\newpage

The simulation results reveal that the total population tends to zero in all the cases, in agreement with Theorem \ref{teomontifin1}. Note that population decreases quickly initially, but after some time it decreases exponentially as $e^{- \mu t}$ and since $\mu$ is small, the decreasing is very slow. It can be verified that with a larger value of $\mu$ the population tends to zero much faster. For the shake of brevity we have not included this simulation in the paper and we have included the most realistic cases. The number of infected people increases and the value of susceptible people decreases faster as $R_0$ increases. The number of hospitalized cases depend on the values of $\gamma$, $\beta$ and $R_0$, and two possible scenarios appear.  First, hospitalized cases depend on infected symptomatic and asymptomatic cases with prevalence of asymptomatic patients (Fig. \ref{my:figf2}a, Fig. \ref{my:figf3}a and \ref{my:figf4}a).  Second, the number of hospitalized patients is always less than the asymptomatic when $R_0\geq 1$ (Fig. \ref{my:figf3}a and Fig. \ref{my:figf4}a). 

\subsubsection{Case $\Lambda=\mu$}

\begin{figure}[h]
\centering
\includegraphics[scale=2.7]{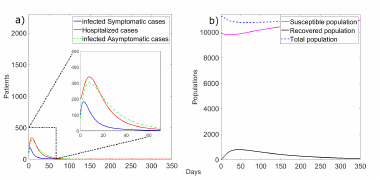}
\caption{\small Simulation of system (\ref{my:eqf1}) for $\Lambda=0.008$, $\mu=0.008$, $ \beta=0.05\ days^{-1}$ and $R_0=0.3496$. a) Infected symptomatic (blue line), hospitalized (red line) and infected asymptomatic (dashed green line)cases b) Susceptible (magenta line), recovered (black line) and total populations (dashed blue line).}
\label{my:figf5}
\end{figure}

\begin{figure}[ht]
\centering
\includegraphics[scale=2.7]{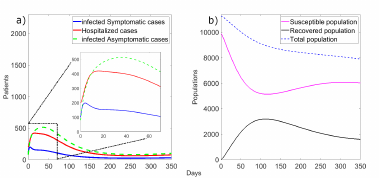}
\caption{\small Simulation of system (\ref{my:eqf1}) for $\Lambda=0.008$, $\mu=0.008$, $ \beta=0.2\ days^{-1}$ and $R_0=1.3983$. a) Infected symptomatic (blue line), hospitalized (red line) and infected asymptomatic (dashed green line) cases b) Susceptible (magenta line), recovered (black line) and total populations (dashed blue line).}
\label{my:figf6}
\end{figure}

\begin{figure}[h]
\centering
\includegraphics[scale=2.7]{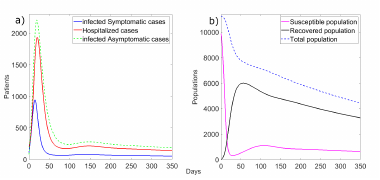}
\caption{\small Simulation of system (\ref{my:eqf1}) for $\Lambda=0.008$, $\mu=0.008$, $ \beta=1\ days^{-1}$ and $R_0=6.9915$. a) Infected symptomatic (blue line), hospitalized (red line) and infected asymptomatic (dashed green line) cases b) Susceptible (magenta line), recovered (black line) and total populations (dashed blue line).}
\label{my:figf7}
\end{figure}

Fig. \ref{my:figf5}a, Fig. \ref{my:figf6}a, Fig. \ref{my:figf7}a show the temporal behaviour of infected symptomatic  (blue line), hospitalized (red line) and infected asymptomatic (dashed green line) cases while in Fig. \ref{my:figf5}b, Fig. \ref{my:figf6}b, Fig. \ref{my:figf7}b the susceptible (magenta line) and recovered (black line) and the total populations (dashed blue line) are shown.
\newpage

\newpage
The simulation results reveal that the solution of system \eqref{my:eqf1} tends to an equilibrium point $(S^*,0,0,0,0,0)$, in all the cases, in agreement with Theorem \ref{otrarevi1}. The value of $S^*$ depends on $R_0$ and decreases as $R_0$ increases. For $R_0>1$, the infected, asymptomatic and hospitalized tend to zero, very slowly, and $S^*$ is zero, in agreement with Theorem \ref{otrarevi1}. The total population $N$ tends to zero faster as $R_0$ increases. The number of hospitalized cases depends on the values of $\gamma$, $\beta$ y $R_0$, showing different possible scenarios.  First, hospitalized cases depend on the symptomatic and asymptomatic infected as shown in Fig. \ref{my:figf5}a, Fig. \ref{my:figf6}a, Fig. \ref{my:figf7}a. Second, the number of hospitalized patients is always lower than the asymptomatic when $R_0\geq 1$, as can be seen in the Fig. \ref{my:figf6}a, Fig. \ref{my:figf7}a.

\subsubsection{Case $\Lambda=\mu=0$}

\begin{figure}[h]
\centering
\includegraphics[scale=2.7]{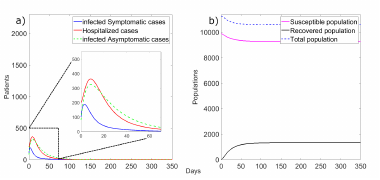}
\caption{\small Simulation of system (\ref{my:eqf1}) for $\Lambda=0$, $\mu=0$, $ \beta=0.05\ days^{-1}$ and $R_0=0.3975$. a) Infected symptomatic (blue line), hospitalized (red line) and infected asymptomatic (dashed green line) cases b) Susceptible (magenta line), recovered (black line) and total populations (dashed blue line).}
\label{my:figf8}
\end{figure}

\begin{figure}[h]
\centering
\includegraphics[scale=2.7]{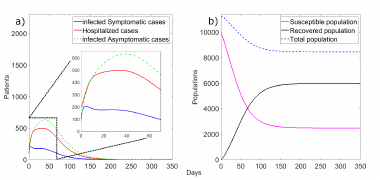}
\caption{\small Simulation of system (\ref{my:eqf1}) for $\Lambda=0$, $\mu=0$, $ \beta=0.2\ days^{-1}$ and $R_0=1.5910$. a) Infected symptomatic (blue line), hospitalized (red line) and infected asymptomatic (dashed green line) cases b) Susceptible (magenta line), recovered (black line) and total populations (dashed blue line).}
\label{my:figf9}
\end{figure}

\begin{figure}[ht]
\centering
\includegraphics[scale=2.7]{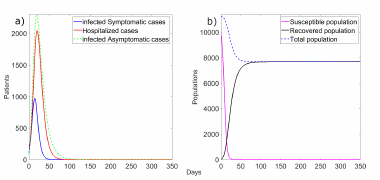}
\caption{\small Simulation of system (\ref{my:eqf1}) for $\Lambda=0$, $\mu=0$, $ \beta=1\ days^{-1}$ and $R_0=7.9500$. a) Infected symptomatic (blue line), hospitalized (red line) and infected asymptomatic (dashed green line) cases. b) Susceptible (magenta line), recovered (black line) and total populations (dashed blue line).}
\label{my:figf10}
\end{figure}

Fig. \ref{my:figf8}a, Fig. \ref{my:figf9}a, Fig. \ref{my:figf10}a show the temporal behaviour of infected symptomatic  (blue line), hospitalized (red line) and infected asymptomatic infected (dashed green line) cases while in Fig. \ref{my:figf8}b, Fig. \ref{my:figf9}b, Fig. \ref{my:figf10}b the susceptible (magenta line) and recovered (black line) and the total populations (dashed blue line) are shown.
\newpage

\newpage

The simulation results reveal that the solution tends in all the cases to an equilibrium point $(S^*,0,0,0,0,R^*)$, in agreement with Theorem \ref{montifininfteo}. For $R_0<1$ the value $S^*$ is not far from the initial value $S_0$ and most of the people do not suffer the infection. The total population decreases due to the infection, but not much. The effect of the disease is moderated. For $R_0 \approx 1$, the values of $S^*$ and $R^*$ are smaller and larger respectively than the ones for $R_0<1$. The population decreases a little more and the effect of the disease is more important than in the previous case.
For $R_0 >1$, $S^*$ is close to zero, which means that almost the whole population gets infected (the recovered people is practically equal to the total population). The population is clearly reduced and the effect of the epidemic is remarkable. The number of hospitalized cases depends on the values of $\gamma$, $\beta$ y $R_0$. These show different possible scenarios. First, hospitalized cases depend on the symptomatic and asymptomatic infected as shown in Fig. \ref{my:figf8}a, Fig. \ref{my:figf9}a, Fig. \ref{my:figf10}a. Second, the number of hospitalized patients is always lower than the asymptomatic when $R_0 \geq 1$, as can be seen in the Fig. \ref{my:figf9}a, Fig. \ref{my:figf10}a. 

\subsection{Model fitting}

To establish the mathematical model, we used official data of accumulated, infected, hospitalized and eliminated cases of COVID-19 in Cuba provided by the\  Ministry\  of\  Public\  Health (MINSAP) that displayed on\  the web\  platform (\href{https://covid19cubadata.github.io}{https://covid19cubadata.github.io}). We choose data reported for 429 days of pandemic from March 12, 2020 to May 15, 2021 for Cuba.

To fit equations \eqref{my:eqf1},  the total population $(N(0))$,  the birth rate ($\Lambda$) and death rate ($\mu$) are calculated from the data reported in the MINSAP statistical yearbook \citep{MINSAP2019}. Values of latent time, $\sigma^{-1}$ (in days),  and the average recovery time of asymptomatic patients ($k^{-1}_1$,  in days) are taken from \cite{lauer2020}.  Initial values of infected ($I(0)$) and hospitalized ($H(0)$) cases are taken from official data. The initial susceptible ($S(0)$) and recovered ($R(0)$) are assumed equal to 99\% of total population and zero, respectively.  Finally, the values of initial asymptomatic patient ($A(0)$), initial exposed patient ($E(0)$), $\Lambda$, $\beta$, $p$, $\delta^{-1}$, $\gamma^{-1}$ and $k^{-1}_2$ are estimated from model.

\begin{table}[h]
\centering
\caption{\small Parameters obtained from model fitting data of Rep\'ublica de Cuba}
\label{my:tabla3}
\begin{tabular}{l c}
\hline
Parameters & Values\\
\hline
Population & 11201549\\
$\Lambda$& 2.68x10$^{-5}$\\
$\mu$&2.67x10$^{-5}$\\
$\beta$&0.2027\\
$p$ & 0.5995\\
$k^{-1}_1$ (days) & 14.0000\\
$\sigma^{-1}$ (days) & 5.1000\\
$k^{-1}_2$ (days) & 10.8342\\
$\delta^{-1}$ (days) & 19.6516\\
$\gamma^{-1}$ (days) & 1.4384 \\
$R_0$&1.3107\\
\hline
\end{tabular}
\end{table}

\begin{figure}[h]
\centering
\includegraphics[scale=4]{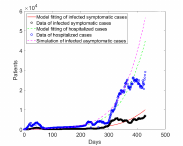}
\caption{\small Fit of system \eqref{my:eqf1} from March 12,  2020 to May 15,  2021. }
\label{my:figf12}
\end{figure}

Fitting the model to the data by a least squares approach, we obtain the parameters shown in Table \ref{my:tabla3}. Fig. \ref{my:figf12} shows experimental data (for symptomatic and hospitalized cases), model fitting (for symptomatic and hospitalized cases) and simulation of infected asymptomatic cases. We can see from this figure that the model fits data of number of infected symptomatic cases but not for hospitalized cases. It is important to point out that all the infected (symptomatic and asymptomatic) cases and suspected individuals of COVID-19 are hospitalized in the Republic of Cuba. Nevertheless, in the model the hospitalized are a fraction of these infected cases.

\newpage
The value of $R_0=1.3107$ obtained from model fitting is close to that reported from experimental data corresponding to March 13, 2020 ($R_0 = 1.3457$).  Both $R_0$ values corroborate our simulations corresponding to the case $R_0\approx 1$ and $\Lambda = \mu$. Values of criteria for model assessment proposed in section \ref{statt} are shown in table \ref{my:tabla4}.

\begin{table}[h]
\centering
\caption{Parameter values of quality of the model fitting}
\label{my:tabla4}
\begin{tabular}{l|cc}
\hline
Parameters & Symptomatic cases & Hospitalized cases\\
\hline
${r^2}^a$ & 0.8533&0.8917\\
${SSE}^b$&1.1751x10$^3$&4.4926x10$^3$\\
${RMSE}^c$  & 1.600x10$^3$&4.4348x10$^3$\\
${RE}^d$  & 0.3825& 0.3287\\
\hline
\end{tabular}
\begin{center}
\footnotesize{${}^a$Goodness of fit, ${}^b$Sum of squares error, ${}^c$Root mean square error, ${}^d$Relative error}
\end{center}
\end{table}

\section{Discussion}

The SEIR model fits well to the reported COVID-19 data with $r^2$ in the range $0.85 - 0.89$. It is plausible that the supplied data, has many fluctuations and likely subject to a large uncertainty. It may be mentioned: 1) fluctuacting behaviours in the number of active and hospitalized cases due to unwanted epidemiological events in the Rep\'ublica de Cuba. These unwanted epidemiological events are due to native transmission and transmission by travellers originating in other countries with high  COVID-19 transmission and new strains of SARS-CoV-2.  2) Introduction of new strains of SARS-CoV-2, as South Africa (variant B.1.551), United Kingdom (variant B.1.1.7), California (variant B.1.429), Brazil (variants B.1.1.28.1 and B.1.2.28.2), and India (variant B.1.617 or delta strain). These strains have higher transmission rates than original Wuhan strain (variant D614G), which prevails in the Rep\'ublica de Cuba from March 2020 up to April 2021. 3) Seven new protocols for confrontation to COVID-19 that have been introduced according to epidemiological situation. 4) Clinical trials (Phses I-III) of Abdala and Soberana-02 vaccine candidates. These reasons explain why model fitting system \eqref{my:eqf1} of experimental data is not totally good in the study period selected, one of limitations of this study. Another limitation of this study is that $\beta$ and $R_0$ are considered constants in our model; nevertheless, epidemiological studies evidence that these two parameters change over time \citep{Fraser2009, Sy2021, Delamater2019, diekmann2010, perasso2018}. 

These two above-mentioned limitations may be solved if this study period is analyzed by section, taking into account the protocol, transmission rate of SARS-CoV-2 strain type, government measures, perception risk of individuals and social responsibility of individuals. For this, a further study is being carried out.

Despite these two limitations, $\beta$, $p$ and $R_0$ values obtained from model fitting to experimental data (Table \ref{my:tabla3}) agree with epidemiological data reported every week to MINSAP, from November 2020 up to the end May 2021. $\beta = 0.2027$ days$^{-1}$ agrees with its $\beta$ values $(0.2 \leq \beta \leq 0.4)$, reported in \citep{Bizet2020, ramireztorres2021}. $R_0 = 1.3107$ belongs to interval $[0.068, 2.85]$, agrees with reported daily by MINSAP.

Simulations and model fitting to experimental data corroborate that the propagation of the COVID-19 disease depends on $R_0$ (Eqs. \eqref{montifin} and \eqref{montijanomodificado35}), which is directly proportional to $\beta$ and $p$ parameters, in agreement with the  definition itself of $R_0$ \citep{Fraser2009, Sy2021, Delamater2019}.  If the time behaviour of $\beta$ (named $\beta(t)$) reported in \citep{ramireztorres2021} is introduced in Eqs. \eqref{montifin} and \eqref{montijanomodificado35}, we would corroborate that $R_0$ depends also on $t$, in agreement with \citep{Fraser2009, Sy2021, Delamater2019, diekmann2010, perasso2018}. 

On the other hand, $R_0$ is directly related to $p$ and its value may contribute to quick COVID-19 propagation  (see Eqs. \eqref{montifin} and \eqref{montijanomodificado35}). The value obtained for $p= 0.5995$ is in agreement with the prevalence of symptomatic cases in the period March 2020 - May 2021. This does not mean that asymptomatic cases become infected later and/or they are capable of infecting other susceptible individuals. This depends on the type of dominant and circulating SARS-CoV-2 strain, strenght of immune system (immunodeficient and immunocompetent patients), and environmental conditions. It should be noted that these asymptomatic patients are not hospitalized because they have no symptoms or are mild. Furthermore, none of them die. From June 2021, it is reported that asymptomatic cases not vaccinated have been infected with the delta variant and some of them with comorbidities have die.

From epidemiological point of view, $p = 0.5995$ means that more than half of the infected cases are detected, indicating that the government policy in the COVID-19 confrontation is correct. Furthermore, this $p$ value agrees with the number of asymptomatic cases reported in the Rep\'ublica de Cuba (55.0\%) in this observation period, in agreement with \citep{Subramanian2021}. 

The values of $k^{-1}_2$, $\delta^{-1}$ and $\gamma^{-1}$ obtained from model fitting of experimental data (Table \ref{my:tabla3}) allow to consider that $\mu$ can be neglected respect to $k_1$, $\gamma$ and $\sigma$ in Eqs. \eqref{montifin} and \eqref{montijanomodificado35}. As $\mu$ is very small, $R_0$ does not change significantly with $\sigma$. This confirms that $R_0$ does not depend on incubation time of any SARS-CoV-2 strain in the host but infected individual that has associated  an average number of infected contacts (definition of $R_0$) \citep{Fraser2009, Sy2021, Delamater2019}. Furthermore, the definition itself of $R_0$ justifies that it does not depend on $\delta$ and $k_2$ either. This is also expected because in our model it is assumed that a hospitalized person does not contribute to generate new infected. Nevertheless, $\delta$ value is very important because it affects directly to the number of deaths due to the infection. In these situations, $R_0$ is dominated by $\beta$ and $p$, as expected. Additionally, results of table \ref{my:tabla3} and $\mu$ very small allow to obtain an expression for $R_0$ when $p=0$ proportional to ${}^\beta/_{k_1}$, this suggest that when $p$ is small the progression of the disease depends on the recovery time average of asymptomatic patients $k_1^{-1}$. On the other hand, if $p=1$ we can obtain an expression for $R_0$ proportional to ${}^\beta/_\gamma$, this suggest that for $p$ large (close to 1) the progression of the disease will depend on the time between symptoms onset to hospiltalization $\gamma^{-1}$. Taking into account that  ${}^1/_{k_1}$ is larger than $1/\gamma$, we  conclude that $p$ large will give smaller values of $R_0$. Consequently, policies tending to get $p$ as large as possible, and making $\gamma$ as large as possible ($k_1$ depends on the disease and can not be modified) will reduce the value of $R_0$.

From temporal behaviour of symptomatic, asymptomatic and hospitalized cases, it can be observed that these take their maximum value depending on the $\beta$, $\gamma$ and $R_0$ values, as shown in the Fig. \ref{my:figf2} to Fig. \ref{my:figf10}. For case $\Lambda<\mu$, the number of Infected cases ($I$) always reaches its minimum value at zero and its quick growth depends strongly on $R_0$. Nevertheless, the number of infected cases for $\Lambda=\mu$ and $\Lambda=\mu=0$ always reaches a minimum value at zero and this depends on the $R_0$ value, being remarkable for $\Lambda=\mu$.

The simulation and fit of proposed model, resulted in that if the time between symptom onset to hospitalization ($\gamma^{-1}$) increases, the average of hospitalization time ($k_2^{-1}$) and average of time that an hospitalized patient died ($\delta$) also increases. This fact, suggests that the earlier the treatment protocol is started, the less time the patient is hospitalized as suggested in \citep{hung2020}. Nevertheless, if we analyse the pandemic in two periods of time the average of time that an hospitalized patient died ($\delta$) should increase due to the development of new treatment protocols \cite{HosseinKhannazer2020} but did not vary much. This may be justified by the presence of new SARS-CoV-2 strains with higher virulence, pathogenicity, invasiveness, and diffusibility, as delta variant.    

The number of symptomatic cases higher than that hospitalized cases is not observed in Rep\'ublica de Cuba. Simulations reveal this unfavourable scenery when the time between symptom onset to hospitalization ($\gamma^{-1}$) increases. This suggest, that the active search of infected can help to control the pandemic.

\section{Conclusion}

We have proposed a modified SEIR mathematical model, that explicitly includes asymptomatic and hospitalized
people, to help the epidemiological study of infectious diseases and we have applied it to the case of COVID-19. We also have include in the model the effect of demography considering that the population grows by a ratio proportional to the total population which leads to solutions whose behaviour is different from those obtained in models where the ratio is considered constant. We have studied the stability of the model and we have  verified these stability results with some numerical simulations. We computed the basic reproduction number corresponding to the model and showed that this number determines the qualitative evolution of the disease. This model was fitted with experimental data from Rep\'ublica de Cuba and we have shown that the model reproduces the evolution of the experimental data. This suggest that the model can be a useful tool. Nevertheless, this model is adequate only for one peak, and this is the reason why research is continuing to obtain an algorithm to improve these results.

\section{Acknowledgement}
This research is supported by the doctoral scholarship from the Universidad de Zaragoza-Santander Universidades program and the projects PID2019-109045GB-C31 of Ministery of Science and Innovation of Spain, Furthermore, We thank Professor PhD. Jes\'us Carnicer for his help in the proof of Lemma 1.

\bibliographystyle{ieeetr}

\bibliography{artico}

\begin{thebibliography}{10}

\bibitem{WHO}
{WHO}, ``worldometers (2021),'' 2021.

\bibitem{li2020}
R.~Li, S.~Pei, B.~Chen, Y.~Song, T.~Zhang, W.~Yang, and J.~Shaman,
  ``Substantial undocumented infection facilitates the rapid dissemination of
  novel coronavirus ({SARS}-{CoV}-2),'' {\em Science}, vol.~368, no.~6490,
  pp.~489--493, 2020.
\newblock Publisher: American Association for the Advancement of Science.

\bibitem{mckibbin2021}
W.~McKibbin and R.~Fernando, ``The global macroeconomic impacts of {COVID}-19:
  {Seven} scenarios,'' {\em Asian Economic Papers}, vol.~20, no.~2, pp.~1--30,
  2021.

\bibitem{Estrada2020}
E.~Estrada, ``Covid-19 and sars-cov-2. modeling the present, looking at the
  future,'' {\em Physics Reports}, vol.~869, pp.~1--51, 2020.
\newblock COVID-19 and SARS-CoV-2. Modeling the present, looking at the future.

\bibitem{Saldana2021}
F.~Salda{\~n}a and J.~X. Velasco-Hern{\'a}ndez, ``Modeling the covid-19
  pandemic: a primer and overview of mathematical epidemiology,'' {\em SeMA
  Journal}, pp.~1--27, 2021.

\bibitem{Brugnano2021}
L.~Brugnano, F.~Iavernaro, and P.~Zanzottera, ``A multiregional extension of
  the sir model, with application to the covid-19 spread in italy,'' {\em
  Mathematical Methods in the Applied Sciences}, vol.~44, no.~6,
  pp.~4414--4427, 2021.

\bibitem{Neves2020}
A.~G. Neves and G.~Guerrero, ``Predicting the evolution of the covid-19
  epidemic with the a-sir model: Lombardy, italy and sao paulo state, brazil,''
  {\em Physica D: Nonlinear Phenomena}, vol.~413, p.~132693, 2020.

\bibitem{Ndairou2020}
F.~Ndaïrou, I.~Area, J.~J. Nieto, and D.~F. Torres, ``Mathematical modeling of
  covid-19 transmission dynamics with a case study of wuhan,'' {\em Chaos,
  Solitons \& Fractals}, vol.~135, p.~109846, 2020.

\bibitem{peng2020}
L.~Peng, W.~Yang, D.~Zhang, C.~Zhuge, and L.~Hong, ``Epidemic analysis of
  {COVID}-19 in {China} by dynamical modeling,'' {\em arXiv preprint
  arXiv:2002.06563}, 2020.

\bibitem{Lin2020}
Q.~Lin, S.~Zhao, D.~Gao, Y.~Lou, S.~Yang, S.~S. Musa, M.~H. Wang, Y.~Cai,
  W.~Wang, and L.~Yang, ``A conceptual model for the coronavirus disease 2019
  ({COVID}-19) outbreak in {Wuhan}, {China} with individual reaction and
  governmental action,'' {\em International journal of infectious diseases},
  vol.~93, pp.~211--216, 2020.
\newblock Publisher: Elsevier.

\bibitem{Ottaviano2022}
S.~Ottaviano, M.~Sensi, and S.~Sottile, ``Global stability of sairs epidemic
  models,'' {\em Nonlinear Analysis: Real World Applications}, vol.~65,
  p.~103501, 2022.

\bibitem{Hu2012}
Z.~Hu, Z.~Teng, and H.~Jiang, ``Stability analysis in a class of discrete sirs
  epidemic models,'' {\em Nonlinear Analysis: Real World Applications},
  vol.~13, no.~5, pp.~2017--2033, 2012.

\bibitem{Saha2020}
S.~Saha, G.~Samanta, and J.~J. Nieto, ``Epidemic model of covid-19 outbreak by
  inducing behavioural response in population,'' {\em Nonlinear dynamics},
  vol.~102, no.~1, pp.~455--487, 2020.

\bibitem{Yang2020}
C.~Yang and J.~Wang, ``A mathematical model for the novel coronavirus epidemic
  in {Wuhan}, {China},'' {\em Mathematical biosciences and engineering : MBE},
  vol.~17, pp.~2708--2724, Mar. 2020.

\bibitem{Shah2021}
K.~Shah, R.~U. Din, W.~Deebani, P.~Kumam, and Z.~Shah, ``On nonlinear classical
  and fractional order dynamical system addressing covid-19,'' {\em Results in
  Physics}, vol.~24, p.~104069, 2021.

\bibitem{Shoaib2021}
M.~Shoaib, M.~A.~Z. Raja, M.~T. Sabir, A.~H. Bukhari, H.~Alrabaiah, Z.~Shah,
  P.~Kumam, and S.~Islam, ``A stochastic numerical analysis based on hybrid
  nar-rbfs networks nonlinear sitr model for novel covid-19 dynamics,'' {\em
  Computer Methods and Programs in Biomedicine}, vol.~202, p.~105973, 2021.

\bibitem{Fraser2009}
C.~Fraser, C.~A. Donnelly, S.~Cauchemez, W.~P. Hanage, M.~D. Van~Kerkhove,
  T.~D. Hollingsworth, J.~Griffin, R.~F. Baggaley, H.~E. Jenkins, E.~J. Lyons,
  T.~Jombart, W.~R. Hinsley, N.~C. Grassly, F.~Balloux, A.~C. Ghani, N.~M.
  Ferguson, A.~Rambaut, O.~G. Pybus, H.~Lopez-Gatell, C.~M. Alpuche-Aranda,
  I.~B. Chapela, E.~P. Zavala, D.~M.~E. Guevara, F.~Checchi, E.~Garcia,
  S.~Hugonnet, C.~Roth, and T.~W. R. P.~A. Collaboration, ``Pandemic potential
  of a strain of influenza a (h1n1): Early findings,'' {\em Science}, vol.~324,
  no.~5934, pp.~1557--1561, 2009.

\bibitem{Sy2021}
K.~T.~L. Sy, L.~F. White, and B.~E. Nichols, ``Population density and basic
  reproductive number of covid-19 across united states counties,'' {\em PLOS
  ONE}, vol.~16, pp.~1--11, 04 2021.

\bibitem{ramireztorres2021}
E.~E. Ramirez-Torres, A.~S. Castaneda, Y.~Rodríguez-Aldana, S.~S. Domínguez,
  L.~V. García, A.~Palú-Orozco, E.~R. Oliveros-Domínguez,
  L.~Zamora-Matamoros, R.~Labrada-Claro, and M.~Cobas-Batista, ``Mathematical
  modeling and forecasting of {COVID}-19: experience in {Santiago} de {Cuba}
  province,'' {\em Revista Mexicana de Física}, vol.~67, no.~1 Jan-Feb,
  pp.~123--136, 2021.

\bibitem{diekmann2010}
O.~Diekmann, J.~A.~P. Heesterbeek, and M.~G. Roberts, ``The construction of
  next-generation matrices for compartmental epidemic models,'' {\em Journal of
  the Royal Society Interface}, vol.~7, no.~47, pp.~873--885, 2010.
\newblock Publisher: The Royal Society.

\bibitem{perasso2018}
A.~Perasso, ``An introduction to the basic reproduction number in mathematical
  epidemiology,'' {\em ESAIM: Proceedings and Surveys}, vol.~62, pp.~123--138,
  2018.
\newblock Publisher: EDP Sciences.

\bibitem{Lakshmikantham1969}
V.~Lakshmikantham and S.~Leela, {\em Differential and Integral Inequalities:
  Theory and Applications: Volume I: Ordinary Differential Equations}.
\newblock Academic press, 1969.

\bibitem{McNabb1986}
A.~McNabb, ``Comparison theorems for differential equations,'' {\em Journal of
  Mathematical Analysis and Applications}, vol.~119, no.~1, pp.~417--428, 1986.

\bibitem{ingber1993}
L.~Ingber, ``Simulated annealing: {Practice} versus theory,'' {\em Mathematical
  and computer modelling}, vol.~18, no.~11, pp.~29--57, 1993.
\newblock Publisher: Elsevier.

\bibitem{Gonzalez2017}
M.~M. González, J.~A.~G. Joa, L.~E.~B. Cabrales, A.~E.~B. Pupo, B.~Schneider,
  S.~Kondakci, H.~M.~C. Ciria, J.~B. Reyes, M.~V. Jarque, and M.~A.~O. Mateus,
  ``Is cancer a pure growth curve or does it follow a kinetics of dynamical
  structural transformation?,'' {\em BMC cancer}, vol.~17, no.~1, pp.~1--14,
  2017.
\newblock Publisher: BioMed Central.

\bibitem{lauer2020}
S.~A. Lauer, K.~H. Grantz, Q.~Bi, F.~K. Jones, Q.~Zheng, H.~R. Meredith, A.~S.
  Azman, N.~G. Reich, and J.~Lessler, ``The incubation period of coronavirus
  disease 2019 ({COVID}-19) from publicly reported confirmed cases: estimation
  and application,'' {\em Annals of internal medicine}, vol.~172, no.~9,
  pp.~577--582, 2020.
\newblock Publisher: American College of Physicians.

\bibitem{VRABEL2015}
R.~VRÁBEL', V.~LIŠKA, and J.~Vaclav, ``Remark on sensitivity of simulated
  solutions of the nonlinear dynamical system to the used numerical method,''
  {\em International Journal of Mathematical Analysis}, vol.~9, no.~43,
  pp.~2749--2754, 2015.

\bibitem{MINSAP2019}
MINSAP, ``{ANUARIO} {ESTADÍSTICO} {DE} {SALUD},'' 2019.

\bibitem{Delamater2019}
P.~L. Delamater, E.~J. Street, T.~F. Leslie, Y.~T. Yang, and K.~H. Jacobsen,
  ``Complexity of the basic reproduction number (r0),'' {\em Emerging
  Infectious Diseases journal}, vol.~25, Jan. 2019.

\bibitem{Bizet2020}
N.~C. Bizet and A.~C.~M. de~Oca, ``Modified sir models for the evolution of
  covid-19,'' {\em Ciencias Matemáticas}, 2020.
\newblock 73-87.

\bibitem{Subramanian2021}
R.~Subramanian, Q.~He, and M.~Pascual, ``Quantifying asymptomatic infection and
  transmission of covid-19 in new york city using observed cases, serology, and
  testing capacity,'' {\em Proceedings of the National Academy of Sciences},
  vol.~118, no.~9, 2021.

\bibitem{hung2020}
I.~F.-N. Hung, K.-C. Lung, E.~Y.-K. Tso, R.~Liu, T.~W.-H. Chung, M.-Y. Chu,
  Y.-Y. Ng, J.~Lo, J.~Chan, A.~R. Tam, H.-P. Shum, V.~Chan, A.~K.-L. Wu, K.-M.
  Sin, W.-S. Leung, W.-L. Law, D.~C. Lung, S.~Sin, P.~Yeung, C.~C.-Y. Yip,
  R.~R. Zhang, A.~Y.-F. Fung, E.~Y.-W. Yan, K.-H. Leung, J.~D. Ip, A.~W.-H.
  Chu, W.-M. Chan, A.~C.-K. Ng, R.~Lee, K.~Fung, A.~Yeung, T.-C. Wu, J.~W.-M.
  Chan, W.-W. Yan, W.-M. Chan, J.~F.-W. Chan, A.~K.-W. Lie, O.~T.-Y. Tsang,
  V.~C.-C. Cheng, T.-L. Que, C.-S. Lau, K.-H. Chan, K.~K.-W. To, and K.-Y.
  Yuen, ``Triple combination of interferon beta-1b, lopinavir–ritonavir, and
  ribavirin in the treatment of patients admitted to hospital with {COVID}-19:
  an open-label, randomised, phase 2 trial,'' {\em The Lancet}, vol.~395,
  pp.~1695--1704, May 2020.
\newblock Publisher: Elsevier.

\bibitem{HosseinKhannazer2020}
N.~Hossein-Khannazer, B.~Shokoohian, A.~Shpichka, H.~A. Aghdaei, P.~Timashev,
  and M.~Vosough, ``Novel therapeutic approaches for treatment of {COVID}-19,''
  {\em Journal of Molecular Medicine}, vol.~98, pp.~789--803, 2020.
\newblock Publisher: Springer.

\end{thebibliography}

\end{document}